\documentclass[sigconf]{acmart}

\usepackage{xspace}
\usepackage{enumitem}
\usepackage{multirow}

\AtBeginDocument{%
  \providecommand\BibTeX{{%
    \normalfont B\kern-0.5em{\scshape i\kern-0.25em b}\kern-0.8em\TeX}}}

\copyrightyear{2022}
\acmYear{2022}
\setcopyright{rightsretained}
\acmConference[UIST '22]{The 35th Annual ACM Symposium on User Interface Software and Technology}{October 29-November 2, 2022}{Bend, OR, USA}
\acmBooktitle{The 35th Annual ACM Symposium on User Interface Software and Technology (UIST '22), October 29-November 2, 2022, Bend, OR, USA}
\acmDOI{10.1145/3526113.3545631}
\acmISBN{978-1-4503-9320-1/22/10}

\newcommand{\sys}{\textsc{FeedLens}\xspace}
\newcommand{\semantic}{\textsc{Semantic Scholar}\xspace}
\newcommand{\control}{\textsc{S2}\xspace}
\newcommand{\aggcount}{$f_{\mathrm{count-over-threshold}}$\xspace}

\begin{document}

\title[FeedLens: Polymorphic Lenses for Personalizing Exploratory Search over Knowledge Graphs]{FeedLens: Polymorphic Lenses for Personalizing \\ Exploratory Search over Knowledge Graphs}

\author{Harmanpreet Kaur}
\authornote{This work was conducted while the author was an intern at the Allen Institute for AI.}
\email{harmank@umich.edu}
\affiliation{%
  \institution{University of Michigan}
  \city{Ann Arbor}
  \state{MI}
  \country{USA}
}

\author{Doug Downey}
\email{dougd@allenai.org}
\affiliation{%
  \institution{Allen Institute for AI}
  \city{Seattle}
  \state{WA}
  \country{USA}
}

\author{Amanpreet Singh}
\email{amanpreets@allenai.org}
\affiliation{%
  \institution{Allen Institute for AI}
  \city{Seattle}
  \state{WA}
  \country{USA}
}

\author{Evie Yu-Yen Cheng}
\email{eviec@allenai.org}
\affiliation{%
  \institution{Allen Institute for AI}
  \city{Seattle}
  \state{WA}
  \country{USA}
}

\author{Daniel S. Weld}
\email{danw@allenai.org}
\affiliation{%
  \institution{Allen Institute for AI \&\\ University of Washington}
  \city{Seattle}
  \state{WA}
  \country{USA}
}

\author{Jonathan Bragg}
\email{jbragg@allenai.org}
\affiliation{%
  \institution{Allen Institute for AI}
  \city{Seattle}
  \state{WA}
  \country{USA}
}

\renewcommand{\shortauthors}{Kaur et al.}

\begin{abstract}
The vast scale and open-ended nature of knowledge graphs (KGs) make exploratory search over them cognitively demanding for users.
We introduce a new technique, {\em polymorphic lenses}, that improves exploratory search over a KG by obtaining new leverage from the existing preference models that KG-based systems maintain for recommending content. The approach is based on a simple but powerful observation: in a KG, preference models can be re-targeted to recommend not only entities of a single base entity type (e.g., papers in the scientific literature KG, products in an e-commerce KG), but also all other types (e.g., authors, conferences, institutions; sellers, buyers). We implement our technique in a novel system,~\sys, which is built over~\semantic, a production system for navigating the scientific literature KG.~\sys reuses the existing preference models on~\semantic---people's curated research feeds---as lenses for exploratory search. \semantic\ users can curate multiple feeds/lenses for different topics of interest, e.g., one for human-centered AI and another for document embeddings. Although these lenses are defined in terms of papers, \sys\ re-purposes them to also guide search over authors, institutions, venues, etc. Our system design is based on feedback from intended users via two pilot surveys ($n=17$ and $n=13$, respectively). We compare~\sys and~\semantic via a third (within-subjects) user study ($n=15$) and find that~\sys increases user engagement while reducing the cognitive effort required to complete a short literature review task. Our qualitative results also highlight people's preference for this more effective exploratory search experience enabled by~\sys.
\end{abstract}




\maketitle

\section{Introduction}

The world's information is shifting from text sources to  knowledge graphs (KGs)~\cite{Suchanek2007YagoAC,Bollacker2008FreebaseAC}; as a result, exploratory search is increasingly performed over these semi-structured sources. Indeed, in 2016, Google's KG included 70 billion facts and answered roughly one-third of the 100 billion monthly searches handled by Google~\cite{gil2022will,vincent_2016}. The size and role of KGs for exploratory search has only grown since then~\cite{heist2020knowledge,fishkin2019less}. For example, a user searching for a new movie to watch might start from the page for their favorite movie, and traverse the KG to see other recent movies where the lead actress has appeared. Likewise, scholars navigate a graph of authors, conferences, and citations in their quest for pertinent literature.
Using existing search tools, exploratory search over KGs is relatively slow and cognitively demanding; its open-ended nature requires an iterative process, wherein people continually acquire knowledge, synthesize it to define what's relevant for their needs, and evaluate the retrieved search results based on a dynamic notion of relevance~\cite{noy2019industry,white2006supporting}. At the same time, many KG-based systems maintain {\em preference models}~\cite{guo2020survey,cao2019unifying}, which they use to recommend new content to users. For example, Netflix and Amazon Prime use view logs to suggest new movies that the user may wish to watch. Similarly, many scholar sites (e.g., Google Scholar) induce a preference model from a user's library contents or explicit `likes,' which they use to recommend new papers to the user.

We present a technique for exploiting these existing personalized preference models to improve exploratory search over {\em all} parts of a KG. Prior work has shown that collecting users' preferences and using them as a ``lens'' to rank and filter search results can make exploratory search more effective.  
For example, SearchLens~\cite{chang2019searchlens} applied user-defined collections of weighted keywords as lenses, which allowed people to view search results for restaurants on Yelp based on their relevance to one or more lenses at a time. We build on this work using two simple but powerful observations: (1) instead of asking for user preferences for a specific search task, we can re-purpose the existing preference models in KG-based applications as ``lenses,'' and (2) a preference model over a base entity type (e.g., movie or paper) can be aggregated over the KG to estimate user preferences for arbitrary {\em collections} of entities of {\em any} type (e.g., movie director, actor; author, paper references). In this way, these extended preference models can be used to rank and summarize all KG content---dramatically improving exploratory search over KGs by making a broad swathe of entities less opaque. We refer to this technique as {\em polymorphic lenses}.

To demonstrate the advantages of polymorphic lenses, we instantiate them in a prototype system that extends \semantic (\control),\footnote{For brevity, we refer to \semantic as \control throughout the paper.} a popular production system for navigating scientific literature, which reifies a complex, multi-entity KG that spans papers, authors, institutions, publication venues, etc. Guided by two pilot surveys ($n=13$ and $17$),
we built~\sys, which leverages users' curated research feeds from \control as preference models. The research feeds rank new papers of interest for the user, based on ratings supplied by the user on previous papers. Users can maintain multiple different feeds to capture different topics or themes of interest.
\sys\ uses these research feeds as polymorphic lenses to rank and summarize not only papers, but also other types of entities in the scientific literature KG, and presents collections of entities across \control with these ranked and summarized entities as new exploratory, navigational end-points. For example, an author's relevance can be defined as the number of their papers that are relevant to a user's research feed; then all relevant authors on a page can be visually signaled as new exploration end-points.

We conducted a within-subjects user study ($n=15$) to evaluate the utility of our new features for exploratory scientific literature search.
Results show that people explored more content when using~\sys without spending significant additional time---user engagement with author-centric features was \textit{20x} greater and system-wide engagement was \textit{4x} greater for~\sys. People reported lower cognitive load (measured via NASA-TLX~\cite{hart1988development}) for~\sys and rated our new features higher on usability (measured via SUS~\cite{brooke1996sus}). The additional exploration did not come at the cost of finding relevant content. \sys also helped people find relevant authors who were not well-known in a field (e.g., students, junior academics). 
In summary, we  make the following contributions:
\begin{itemize}
    \item We present \textit{polymorphic lenses}, a simple but powerful generalization of ``lenses'' from prior work~\cite{chang2019searchlens}, that re-purpose existing user preference models to serve as lenses across {\em all} types of entities and collections in a knowledge graph. We show how polymorphic lenses enable new ranking and summarization features that improve exploratory search.

    \item We present a system,~\sys, that instantiates our approach for exploratory search in the scientific literature domain. We evaluate {\em summary embeddings}, an optimization that helps provide real-time performance when implementing polymorphic lenses for complex multi-entity knowledge graphs with billions of entities. 
    
    \item We present a quantitative evaluation showing that~\sys is comparable to current search practices for finding relevant, diverse, and novel content, and it does so with reduced cognitive load and higher user engagement. 
    
    \item We perform a qualitative comparison, which shows that \sys provides additional benefits for the sensemaking, filtering, and organizing aspects of exploratory search compared to existing literature review practices. 
\end{itemize}

\section{Related Work}
Our work adds to the extensive literature on capturing user interests to improve the exploratory search experience for knowledge graph-based applications. Exploratory search is an open-ended investigation of a  conceptual area (e.g., part of a knowledge graph) where users' information needs emerge over time as they learn and investigate new information~\cite{white2006supporting,marchionini2006exploratory,ruotsalo2013supporting}. As such, a key challenge of exploratory search is the dynamic and contextual nature of the search query. The Information Retrieval and Human-Computer Interaction communities have jointly proposed several high-level goals for an exploratory search system~\cite{white2009exploratory}. Most prominent amongst these is to include the human in the search approach (e.g., in the form of people's dynamic needs, skills, and social resources~\cite{marchionini2006exploratory}).

Prior work on capturing human feedback for exploratory search can be broadly categorized as either being applicable to a single search session or maintaining user preference models over time. The former is usually the case with faceted search, where people filter their query-based search results based on automatically extracted conceptual dimensions about the entities in a knowledge graph (e.g., authors, publication venue, and publication year for scientific literature)~\cite{yee2003faceted,o2010tweetmotif}. While faceted search extracts dimensions based on entities in a knowledge graph, clustering mechanisms apply the same principle to the results of a search query. These clusters can similarly be used to filter results based on user feedback~\cite{hearst2006clustering}. Sciascio et al. apply this approach coupled with interactive features that allow users to explore different combinations of keywords and the corresponding re-sorted list of results on-the-fly~\cite{di2016rank,sciascio2019roadmap}. \citet{peltonen2017topic} apply this clustering approach in visual manner, proposing a radial layout for visualizing both relevance and topic similarity in search results. While these approaches that are applied to single search sessions have the benefit of no start-up costs, they do not retain and reflect user interests or knowledge over time. Both of these types of capturing human feedback for exploratory search rely on unsupervised approaches to generate filters or relevant organization for search results per session.

On the other hand, personalized exploratory search systems collect user feedback over time and shape the search experience accordingly. This can be done in implicit ways (e.g., via search logs or user interactions such as clicks~\cite{teevan2005personalizing}) which avoids start-up costs, though privacy can be a concern~\cite{chellappa2005personalization,toch2012personalization}. \citet{palani2021conotate} extend this approach using richer contextual information by mining the notes people take during the search process. They find patterns of knowledge in these notes to support deeper dives into specific topics, and identify gaps in knowledge to suggest query formulations that support exploration. In a similar vein of relying on signals from groups of people, prior work has also proposed approaches for collecting and applying community-generated tags to filter and organize search results~\cite{sen2006tagging}. These tags essentially serve as socially-grounded navigational signposts, and have been successfully instantiated as such in some systems (e.g.,~\cite{kammerer2009signpost,schwarz2011augmenting}).

\begin{figure*}[tb]
    \centering
    \includegraphics[width=0.8\textwidth]{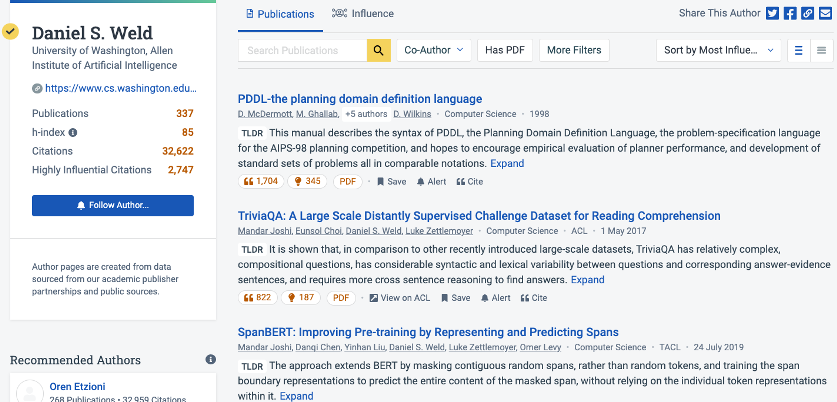}
    \caption{The baseline \semantic interface for an author's homepage. It includes functionality for viewing and searching within an author's publications; sorting them by recency, citations-based metrics, most influential papers; viewing relevant, recommended authors; and adding email alerts for author updates. Compare with our \sys interface in Figure~\ref{fig:system_ui}.}
    \label{fig:scholar_ahp}
\end{figure*}

More related to our work is the subset of personalized exploratory search systems which explicitly capture user feedback, giving them full control over defining their interests and updating these over time. Users can iteratively provide feedback on automatically extracted keywords and metadata (e.g.,~\cite{Glowacka2013DirectingES,ruotsalo2013supporting}) or collections of search results (e.g.,~\cite{baldonado1997sensemaker}). \citet{chang2019searchlens} extend this form of explicit feedback to user-defined lenses, which are reusable collections of weighted keywords that can be applied individually or collectively to match different configurations of a user's needs. Our work builds on this in two main respects: we reuse existing preference models that are common on search and recommender systems---thus saving work for the user---and apply them as {\em polymorphic} lenses that aid exploratory search over multiple types of entities in a KG (i.e., not just the type for which they were originally trained).

Our work is also grounded in prior work on visualizing search results using different approaches for sensemaking~\cite{bernard2013motionexplorer}. Two well-established models that have been applied in the search context are: (1) ``focus+context'' navigation, where the visual representation highlights the primary object(s) of interest (e.g., top 5 search results or papers under a category), but also provides the surrounding context (e.g., the remaining search results or relationship between all categories) for continued exploration~\cite{fox2006exploring,luciani2018details}; and (2) ``overview+details'' navigation, where the visual representation provides an overview of the search results and allows users to choose which ones to view in detail~\cite{bernard2013motionexplorer}. \citet{shneiderman2003eyes} adds a task-centric aspect to the latter, making it ``overview first, zoom and filter, details-on-demand,'' where people gain an overview of the entire collection, zoom in on items of interest and filter out the others, and select items for which they want more details. This approach also includes understanding relationships among items by easily switching between the overview and detailed visuals, refining user interest models based on the new information learned or by extracting relevant items from the list of search results. We follow Shneiderman's model since it lends itself well to rich and varied sets of information. This visualization approach for exploratory search has been used in systems like Atlasify, where spatial reference systems are used to indicate semantic relatedness~\cite{hecht2012explanatory}. In practice, our ``lens'' metaphor can be described as a novel preference-based instantiation of a reference system described in~\cite{hecht2012explanatory}.

With the inherent need for sensemaking in exploratory search, there is also a need for explainability of the search results. Recommender systems rely on black-box models or algorithms. In some cases (e.g., social media news feeds), people do not want an explanation for the curated content that is shown to them~\cite{eslami2015always}. However, in a task-based setting like scientific literature search, providing some form of explanation for recommendations can improve the user experience~\cite{radensky2022exploring}. Our system's visuals and the explanations provided about the data underlying these visuals are designed to serve this purpose. While prior work has provided explanations for a single base entity (e.g., papers in a scientific literature recommender~\cite{lee2020explanationbased}), our polymorphic lenses extend this capability by explaining recommendations for additional types and collections of entities in a KG (e.g., authors, venues, and institutions).

Research on exploratory search encompasses a variety of approaches for sensemaking, visualization, personalization, knowledge graph traversal, and several other topics that are research domains in their own right. We have presented a very small subset of this work here---work that directly inspired our technique and system design. For a more comprehensive review of this space, please refer to the synthesis by \citet{white2009exploratory}. 

\section{Setup and Definitions}
We consider the context where a user is performing exploratory search over a knowledge graph---a directed graph whose vertices denote typed entities, which have attributes. Directed edges connect entities and are labeled with associated relations. For example, in the domain of scientific literature, entities include papers, authors, venues, institutions, etc. Attributes include name and institution for authors, and date for papers. The {\sc writtenBy} relation connects papers to authors, and the {\sc cites} relation connects a paper to ones that it references. These attributes and relations induce {\em content lists} of entities (e.g., all papers written by an author or those citing a paper), which a Web application may display, along with navigational links to other lists. Figure~\ref{fig:scholar_ahp} shows a content list of all papers written by an author. Other domains have similar structure, e.g., a KG for movies connects the actor and movie entities using the \textsc{appearsIn} relation, and the movie and production company entities using the \textsc{filmedUnder} relation. Each of these entities has its own attributes, e.g., genre, year, budget for movies, and years active, nominations, awards for actors. Content lists comprising these entities could take the form of all movies that an actor has appeared in or those directed by a production company. 

Exploratory search over a KG is often open-ended and can quickly become challenging and cumbersome. We now describe two resources from prior work, which are integral to our technique introduced in the subsections that follow:
\begin{enumerate}
    \item \textit{User preference models.} Commonly applied to KGs, preference models are structured representations of a user's interests that are used in personalized search and recommender systems to organize content lists based on their relevance to the user~\cite{jawaheer2014modeling}. In our context, they can be thought of as an arbitrary {\em regression model} that scores the relevance of an entity in the KG. Preference models are typically trained over and applied to a common {\em base entity type} in the KG, such as a paper, movie, song, or product.
    \item \textit{Lenses for exploring a set of entities.} Prior work has used ``lenses''---reusable collections of weighted keywords that represent a user's interests~\cite{chang2019searchlens}---to enable exploration of a set of entities using custom, user-defined criteria. One or more lenses can be applied to an entity to produce relevance scores that represent the user's interest in the entity in the context of each lens. \citet{chang2019searchlens} built SearchLens, a standalone system that allows for exploration of a set of (unlinked) restaurant entities and associated text (customer reviews), prioritized by user-defined lenses. For example, using SearchLens for restaurant recommendations, one could create an ``Outdoor Seating'' lens with different weights for keywords such as patio, balcony, dog-friendly, and receive recommendations ranked by Outdoor Seating.
\end{enumerate}

\begin{figure*}[]
    \centering
    \includegraphics[width=\textwidth]{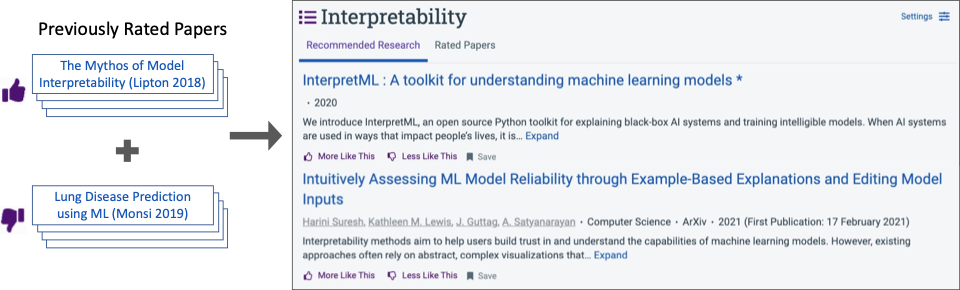}
    \caption{\semantic allows people to annotate papers as ``More'' or ``Less Like This'' and then uses this labeled data to train a ranking function which powers paper recommendations. Here you can see two recommendations for a user's ``Interpretability'' feed. By treating these learned ranking functions as polymorphic lenses, \sys can personalize many aspects of the \semantic UI, as shown in Figures~\ref{fig:CHI} and \ref{fig:system_ui}.}
    \label{fig:scholar_feed_recs}
\end{figure*}

\section{\semantic: The Baseline}
\semantic (\control) is a production system for literature search that relies on KGs. The system provides search functionality, and the search results are linked to detailed views of papers (e.g., relevant figures, citations, references, related papers) and authors (e.g., all of the authors' papers, search functionality within the authors' papers, similar recommended authors, etc.). Figure~\ref{fig:scholar_ahp} shows the detailed author view (i.e., an author's homepage) from \control. The author homepage has links to all of the author's papers, sorted by most influential ones. Users can change the sort order to be based on recency or citations-related metrics including total citation count; citations per year; and most influential citations, calculated based on a ML model’s classification of the citation and its surrounding text as incidental (e.g., as a related work, comparison) or important (e.g., direct use, extension)~\cite{valenzuela2015identifying}. Other features include search within the author's publications, recommended authors, options to view and save individual papers by the author to the user's library, etc. 

\control also allows people to create research {\em feeds}---a stream of newly-published papers recommended based on a machine-learned preference model. To set up a feed, the user annotates a few papers as liked (``More like this'') or disliked and then \control automatically trains a model to suggest similar papers. For example, consider a research feed on ``Interpretability'' curated by a user to include human-centered interpretability work (Figure~\ref{fig:scholar_feed_recs}). Based on the annotated papers,~\control provides recommendations for other papers that might be relevant and allows users to annotate these as well, as further signals to the research feed-based recommender. \control includes necessary features for users to set up and manage the research feed: users can examine and edit their lists of annotated papers for each feed, rename their feeds, and use search or browsing to find and add new papers to their feeds. 

These feeds are a \textit{user preference model}, as defined in the previous section, and we extend them to be used as \textit{lenses}, as described in our system section below.

\section{\sys}
\sys extends the baseline \semantic production system by taking advantage of the existing user preference models in~\control (i.e., the research feeds) and reusing them as ``polymorphic lenses,'' to view and understand other entities (e.g., authors, conferences) in the context of the base entity (research papers). Next, we first describe our design desiderata obtained via pilot studies, followed by details on our polymorphic lenses approach, our frontend interface and the backend implementation.

\subsection{Pilot Surveys}\label{sec:phase1}
\sys is grounded in design feedback from prior work (e.g.,~\cite{norman2013design,raskin2000humane}) as well as the scientific literature application-specific guidelines from our pilot studies presented below.\footnote{Details on our pilot studies are included in the supplementary material.} We conducted two design pilots, one each with~\control system designers and researchers ($n=17)$, and current~\control users ($n=13$). These surveys helped us gather the preferred feature designs (presented in Section~\ref{sec:system}) from both people who work on~\control (which helped constrain our mock-ups to ones that could feasibly be added to~\control) and people who use it for literature search. Based on a thematic analysis of the open-ended feedback on preferred designs---conducted as open coding followed by axial coding~\cite{corbin1990grounded}---we derived the following design guidelines for our system:
\begin{enumerate}[label=\textbf{D\arabic*.}]
    \item \textbf{Support quick perusal.} When displaying information for multiple types of entities at the same time, balancing informativeness and (visual) information overload is critical. People preferred visualizations that supported quick perusal of information.
    \item \textbf{Embed in existing design.} The new relevance information must be embedded in the existing interface design and not distract from information already on the page.
    \item
    \textbf{Build trust with supporting details.} While quick perusal of relevance was critical, participants also wanted the option to dig deeper if they were so inclined. They wanted additional information on how and why relevance was calculated, to build trust in the system.
\end{enumerate}

\begin{figure*}[t]
    \centering
    \includegraphics[width=0.8\textwidth]{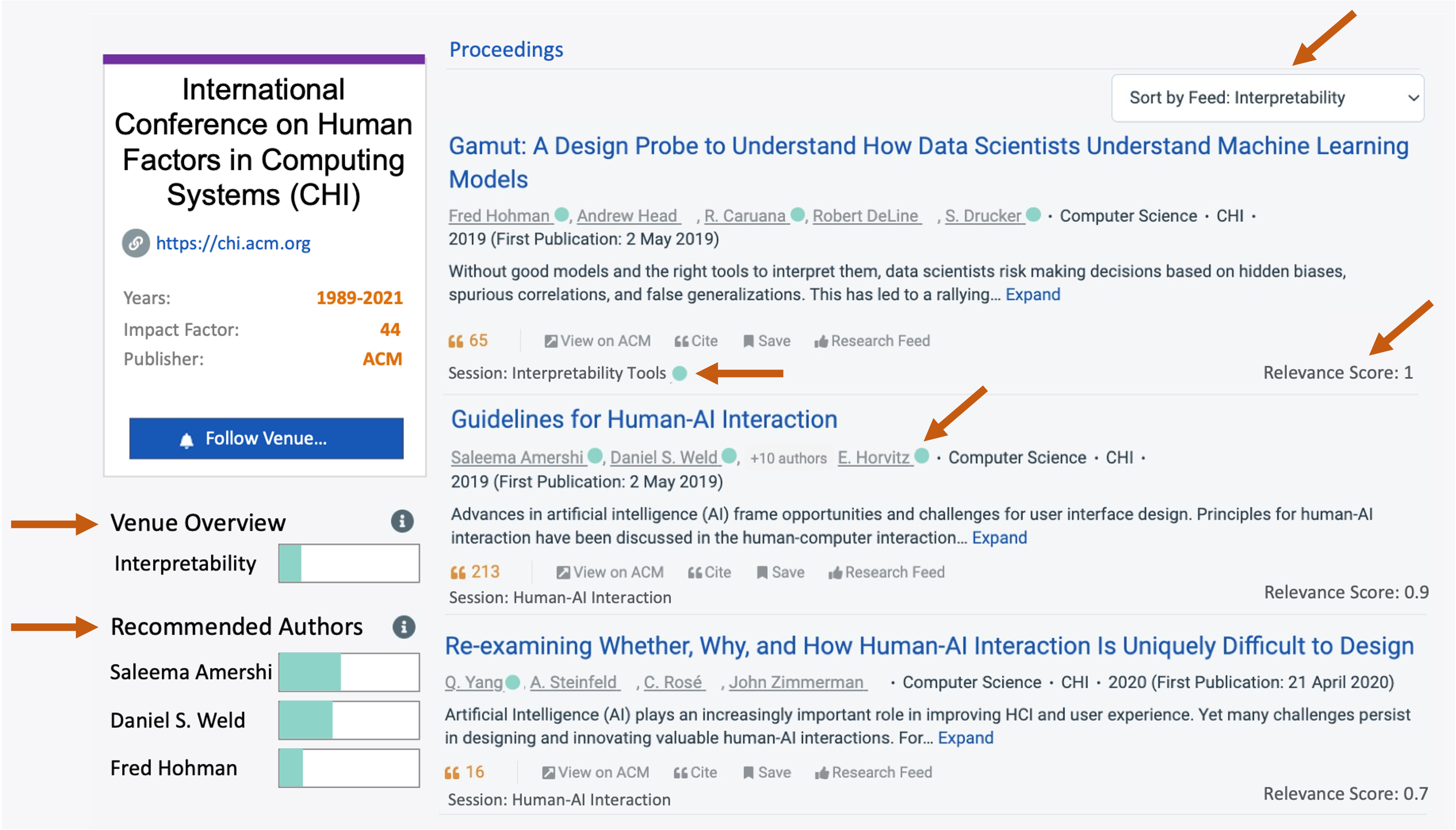}
    \caption{Visualizing the conference CHI 2019, with a lens defined for ``Interpretability.'' The system scores each paper's relevance to the lens (displayed with the ``Relevance Score:'' prefix at the bottom right of each paper item). The papers are ranked by this relevance score (indicated by the selected `Sort by' option, upper right). By applying ``Interpretability'' as a polymorphic lens, we can also score and rank authors (as shown in the ``Recommended Authors'' section in the bottom left). There are also several visualization options for summaries, using text for session relevance scores (right), bars for recommended authors (bottom left), and green circles next to authors and sessions deemed highly relevant to the lens.}
    \label{fig:CHI}
\end{figure*}

\subsection{Polymorphic Lenses}\label{sec:polymorphic}
\sys uses existing user preference models in~\control (i.e., the research feeds based on base paper entities) as ``polymorphic lenses'' to score, rank, and summarize other types and lists of entities (e.g., authors, conferences, institutions). The benefit of our polymorphic lenses approach lies in its simplicity and generality. Preference models trained over a base entity type are ubiquitous---they are used for ranking and recommendation purposes in many applications (e.g., for papers or movies). Our approach simply extends the use of this underutilized resource from the base entity to multiple entity types and collections in a KG. Moreover, since most web applications rely on preference models, our approach can be seamlessly integrated in existing interfaces, with ranking and summarizing affordances.

\subsubsection{Polymorphic Lenses: Formal Definitions}
\label{sec:ranking}
Polymorphic lenses extend a lens defined on a base entity type to serve as a lens for any other related type or collection of entities in a KG. Roughly speaking, a polymorphic lens estimates a user's preferences for a non-base entity (e.g., an author, a conference) by aggregating their estimated preferences over the related base entities (e.g., the papers written by an author, the papers that appear in the conference).

Formally, let $P_i(b) \rightarrow [0,1]$ denote the preference model for lens $i$, defined for any given base entity $b$. For any other entity type $T$ that is connected to the base type by a relation  $R$ in the KG, we define the polymorphic lens $P^T_i(t)=f_{\mathrm{a}}(\{P_i(b) \mid b\in R(t, b)\})$, where $t$ is an entity of type $T$ and $f_{\mathrm{a}}$ is an aggregation function over base entity preference scores.%
\footnote{Note that polymorphic lenses can be applied recursively to entity types $T_2$ that are indirectly connected to the base type via an intermediate entity type $T_1$, as $P^{T_2}_i(t_2)=f_a(\{P^{T_1}_i(t_1) \mid t_1 \in R_2(t_2, t_1) \})$, where $t_1\in T_1$, $t_2\in T_2$, and $R_2$ is a higher relation.}
For example, ~\sys uses $f_{\mathrm{a}}=f_{\mathrm{count-over-threshold}}$, which estimates the {\em number} of preferred base entities $b$ (i.e., those with $P_i(b)$ above some threshold) that are in a relation with $t$.

A polymorphic lens $P^T_i(t)$ can be used to {\em rank} entities of type $T$ in terms of estimated preference to the user. They can also be used to {\em summarize} an individual entity in terms of its scores across potentially multiple lenses. We illustrate these capabilities in the context of conference exploration in the following subsection.

\subsubsection{Polymorphic Lenses: Example with Conference Entities}
Figure~\ref{fig:CHI} shows a mock-up of a web page for the CHI conference entity, grounded in our design desiderata (Section~\ref{sec:phase1}). Applying the user's ``Interpretability'' research feed as a lens, the system scores each paper's relevance to that concept (displayed with the ``Relevance Score:'' prefix at the bottom right of each paper item in the list). These papers can then be sorted according to this preference model for quick perusal (indicated by the selected ``Sort by'' option, upper right); the figure shows paper relevance scores in descending order, with the most relevant papers at the top. Since this is a polymorphic lens, it can also score and rank authors, as shown in the ``Recommended Authors'' section in the bottom left (with the %
$f_{\mathrm{count-over-threshold}}$ relevance scores indicated using bars rather than text). In this way, ranking entities with lenses enables rapid exploration and discovery of relevant entities within any list. 

\begin{figure*}[t]
    \centering
    \includegraphics[width=0.9\textwidth]{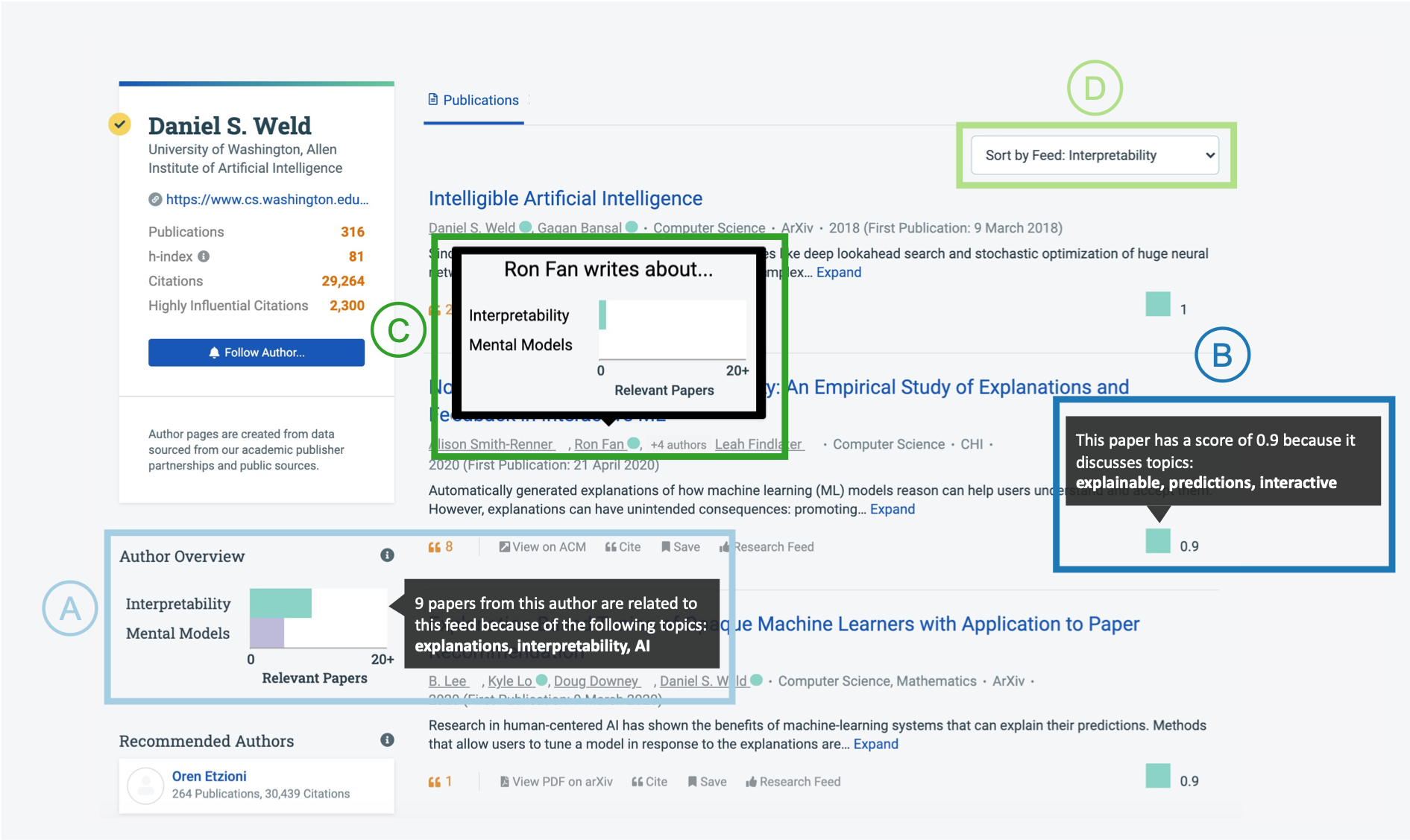}
    \caption{\sys features: (A) Author relevance overview and explanation available on hover, (B) Paper relevance score and explanation on hover, (C) Author recommendations as green circles with author relevance overview on hover, and (D) Lens-based default sort.}
    \label{fig:system_ui}
\end{figure*}

The bars in the recommended authors section also illustrate the new summarization capability made possible by polymorphic lenses.  At a glance, the user can see how related each recommended author is to the Interpretability feed, along with a similar measure for the conference as a whole (shown above the recommended authors). Other authors that are relevant to the feed are called out in the paper list, with green circles (see Section \ref{author_recs}). These features succinctly summarize the relevance of the entities to the user's feed, information that would otherwise require the user to navigate to and inspect each entity individually. Using multiple lenses simultaneously can further enrich the summaries (Figure \ref{fig:system_ui}).\looseness=-1

\subsection{Frontend Interface Design}\label{sec:system}
Based on the design desiderata from our pilot studies, we implemented four new features that apply a user's research feeds as polymorphic lenses for papers (the base entity in a scientific literature KG) and authors (an additional entity type). Although our polymorphic lenses approach can be applied to all types of entities in this KG, we implement these ideas for  authors  for two reasons: (1) to ensure consistency with the functionality afforded by the baseline~\control system (for our within-subjects user study, presented in Section~\ref{sec:study}); and (2) to test the feasibility of our approach in a concrete, yet development cost-sensitive way---an acceptable format for systems research and contributions~\cite{fogarty2017code,buxton2010sketching}.

The features described below are implemented on top of the existing features of~\control. Prior work suggests that this approach of ``piggyback prototyping'' can be particularly beneficial when working with large-scale production systems~\cite{grevet2015piggyback}. In line with that,~\sys simply aims to make the author entities less opaque via the new features. Users can still judge author relevance using~\control alone by clicking on an author entity, which loads an author homepage (Figure~\ref{fig:scholar_ahp}) that includes helpful features like searching within an author's papers and viewing recommended authors. All design choices during implementation described below (e.g., choice of x-axis labels in a bar chart, thresholds for displaying relevant items) were made based on pilot polling with 3--5 people.

\subsubsection{Author Overview and Explanations. }\label{author_overview} 
The polymorphic lens defines an author's relevance to the lens using \aggcount, the count of their papers that are above the relevance threshold.  This is visualized (summarized) using a bar chart (Figure~\ref{fig:system_ui}, feature A), which was the preferred format selected by~\control users in pilot surveys. The bar chart visualization provides a succinct summary, making it easy to quickly peruse this information and determine whether an author is relevant (design guideline D1). The visualization is displayed under the author details card, thus embedding it in the left pane which presents high-level details about an author on~\control (D2). People can also view the raw relevant paper counts as well as a high-level explanation for why the author is deemed relevant (D3), when they hover over individual bars (Figure~\ref{fig:system_ui}, feature A). Details on how these explanations are generated are included in the backend implementation section below.

Using relevant paper counts to define author relevance was a deliberate design choice meant to balance people's preference for seeing numerical values for building trust (D3) and ease of understanding what the value represents (D1). We sampled several other values including proportion of relevant papers per author, and log transformations of relevance counts and proportions---paper counts met both design considerations compared to other options.

To provide a single range of publication counts for helpful comparison between authors, we selected the range [0--20+] relevant papers for the x-axis of the author overview chart. Prior to making this change, we used a larger range to accommodate the high variance in number of publications across authors from different domains, but received feedback that it was difficult to distinguish between authors with fewer relevant publications.
Improving the readability of smaller counts also helps to counter the rich-get-richer effects common in KG-based communities, by de-emphasizing the differences between junior and senior authors (who may have hundreds of publications and would otherwise have much larger bars).

\subsubsection{Paper Relevance Scores and Explanations}\label{paper_squares}
Paper relevance is established via the presence of a visual signal---Figure~\ref{fig:system_ui}-feature B shows our signal, a square that matches the color assigned to the lens in the author overview chart. When a paper's relevance score is above a threshold of 0.5 (the classifier decision boundary) on a scale of [0--1], the paper row displays the square and the score. Displaying positive relevance values in the range [0.5--1] causes users to interpret that the paper is relevant, which we did not find to be the case with values less than 0.5. The presence of the visualization itself indicates that the paper is important for the lens; a quick perusal is all that is needed to internalize this (D1). 

Paper relevance is tricky to visualize in part due to the paper row already being compact and full of information. Figures~\ref{fig:scholar_ahp}--\ref{fig:scholar_feed_recs} all show the high information content in a paper row in the~\control interface. Yet, people in the pilot studies still preferred designs where paper relevance information was embedded in the existing paper row (D2). We ultimately decided on the visualization with squares and scores since it also balanced the other design guidelines: it allowed for a quick perusal of relevance (D1), and also provided a signal for building trust in the underlying recommender (D3) in the form of explanations for the paper's relevance score.

\subsubsection{Author Recommendations}\label{author_recs}
Finding a balance between quick perusal of information (D1) and having access to information for building trust (D3) is far more challenging for lists of entities (e.g., all authors across all papers in a content list, like a paper search results page)---there is far more information available for a list of entities, which can clutter an interface and lead to information overload. We circumvent this issue by using the summarization function of polymorphic lenses and adapting the ``overview first, zoom and filter, then details-on-demand''~\cite{shneiderman2003eyes} model of navigation as our \textit{progressive summarization} approach, by: (1) enabling streamlined browsing by visually highlighting the relevant authors from the list of all authors on the page, and (2) providing access to additional information for this subset of authors on-demand.

Figure~\ref{fig:system_ui}-feature C shows an example of the green circles that we embedded next to the author links on a page to indicate relevance (D2). This feature (termed `Author Recommendation Dots' to users) is designed to draw attention to the subset of relevant authors on a page, with character-sized filled markers chosen to make their presence and color salient at a quick glance (D1). We add information that explains the recommendation in the form of the author overview charts (Section~\ref{author_overview}) which can be accessed by hovering over the author name or the green circle attached to it (Figure~\ref{fig:system_ui}, feature C) (D3). This feature also supports efficient high-level comparisons between the recommended authors. For example, a user could hover over two recommended author names in quick succession and select the author with more/fewer relevant papers to explore further by navigating to their author homepage.

Naturally, the selection of relevant authors requires some threshold to avoid recommending too many authors and defeating the purpose of a streamlined browsing experience. We selected five relevant papers as this threshold after pilot testing and polling with a range of values. That is, an author is recommended using the green circles if they have at least five papers that are relevant to a lens. While this is effective in narrowing down the list of recommended authors on a page, we did not want to bias against authors who were new to a research area (e.g., graduate students, researchers who shifted their focus and recently started working in that domain) who might not have as many relevant papers. To avoid this and encourage exploration, we also recommend a random subset of 50\% of the authors on the page who have [1--5) relevant papers. This, again, is intended to help counter the rich-get-richer phenomenon and was included in the pilot testing.

\subsubsection{Feed-based Sorting}
We enable streamlined browsing of relevant papers on a page (D1) by presenting paper lists ordered by individual paper relevance scores (Figure~\ref{fig:system_ui}, feature D), scored and ranked as discussed in Section \ref{sec:ranking}. This sort order is visually signaled by the individual paper squares and scores feature (Section~\ref{paper_squares}), already embedded in the existing interface (D2). The feed-based sort option is selected by default for paper lists on author homepages to show most relevant papers first; people also have access to other sort options using the same interface. All papers with squares are relevant, and people can further explore why, using the numerical values included with the squares as well as the explanations that can be accessed by hovering over the square (D3).

\subsection{Backend Implementation \&\ Scalability}\label{sec:backend}
Our backend preference model is an ensemble regression model that scores each paper by averaging the scores from two different linear Support Vector Machine (SVM) models. One of the SVMs uses textual (unigram and bigram) features with tf-idf normalization, following the public arxiv-sanity research paper recommender\footnote{\url{http://www.arxiv-sanity.com/}}.  The other model represents each paper with SPECTER embeddings, which are produced by a pre-trained language model (SciBERT \cite{Beltagy2019SciBERT}) fine-tuned on a citation-prediction objective and have been shown to be effective for recommendation \cite{specter2020cohan}. Both models take the title and abstract of each paper as input. The models are trained on the binary paper ratings provided by the user, along with a small set of randomly drawn papers from the corpus that serve as "pseudo-negative" ratings for regularization. The model is also re-trained before application each time a user updates their paper ratings. The SVMs are classifiers that produce decision scores, which we linearly map to the $[0,1]$ range of the preference model.  Specifically, we map each ensemble score output $s$ to a preference value of $\max(\min((s-\tau)\gamma+0.5, 1.0), 0.0)$, where $\tau$ is the decision-score threshold that maximizes accuracy in three-fold cross validation on the user's papers, and $\gamma$ is a constant (for all users), set heuristically.

To produce the explanations, we use LIMEADE \cite{lee2020explanationbased}, which follows LIME \cite{lime2016} in selecting the most informative features from a local linear approximation to the classifier (in our case, we simply use the global linear tf-idf model from our ensemble).  LIMEADE diversifies the explanations by aggregating scores across terms that share the same stems, using the Porter stemmer.

Scalability is a significant challenge in our application.  Many different knowledge graph elements of different types may need to be ranked or summarized on a given page.  Further, summarizing even a single non-base entity can be expensive, as in the case of a prolific author---naively, computing the summary requires scoring each of the author's papers.  For example, in a list with $p$ papers, $a$ authors per paper, and $w$ papers written by each of those authors, the number of base entities to be scored grows as the {\em product} of these three values, i.e. $O(paw)$, which can be large.  In practice, we estimate that \sys\ scores more than 200 papers per pageview, on average.  Lastly, keeping the lens up-to-date with user preferences requires {\em re-training} the model on the latest preference data before applying the lens.

We address the scalability challenge in our prototype in several ways.  We use efficient (linear) models, and we also exploit the fact that polymorphic lenses only require a preference model defined over a base entity type (as discussed in Section \ref{sec:polymorphic}).  This means that we can batch all the base entities that need to be scored for a page into a single request, so that the classifier only needs to be trained once on the most recent preference data.

Our user studies suggested that further improvements in scalability would be needed for large-scale deployments, so we also developed and evaluated a more efficient approach for summarization in polymorphic lenses, to be incorporated in \sys\ in future work.  This approach avoids the expensive step of exhaustively executing the recommender for every related base entity when summarizing a given non-base entity.  Instead, we pre-cluster the base entity representations into a succinct set of cluster centroids, called {\em summary embeddings}, and only run the recommender on this smaller set.  In our application, we cluster an author's paper embeddings, and then at query time given a feed run only the cluster centroids through the recommender, and we estimate the total number of related papers as the summed size of the clusters that have related centroids.

Our experiments show that using summary embeddings can improve runtime efficiency at a small cost to accuracy.  To analyze the effects, we constructed a data set using 346 real feeds from our system. In order to obtain a challenging and balanced workload, we took the top 500 recommendations from each feed, and selected from them 10 "positive" examples (authors with nonzero relevance to the feed), 10 "hard negative" examples (authors within the top 500 papers but with zero relevance to the feed).  To this set we added 10 "easy negatives" randomly drawn from the corpus, for a total of 30 evaluation authors per feed. We experimented with K-means and greedy agglomerative clustering to create summary embeddings, and found the former to perform somewhat better.  Our results on a test set (20\% of the data) are shown in \autoref{tab:author_embed}.  When using for example K=$\sqrt{n}$ summary embeddings instead of an author's $n$ paper embeddings, we see a factor of 12 speedup at the cost of a root mean squared error (RMSE) of about 4.  For interpretability, we also report how often the predicted count $p$ is approximately within a factor of two of the true count $t$, specifically when $1/2 \le (t+1)/(p+1) \le 2$.  With $K=\sqrt{n}$ this is true 88.5\% of the time. Increasing the number of clusters decreases the speedup but also reduces error (with K=$n$ clusters, one for each paper, the method reduces to the original exhaustive approach).

\begin{table}[]
\resizebox{\columnwidth}{!}{
\begin{tabular}{@{}cccc@{}}
\toprule
Experiment                     & RMSE  & \% Within Factor of \textasciitilde2*$t$ &    Speedup\\ \midrule
Baseline - Mean Relevant Count  & 22.12 & 14.90  &   - \\
Single Cluster per author         & 21.57 & 80.62    &   -\\
K-Means (K=0.25$\sqrt{n}$)              & 7.85  &   82.14   &   47$x$\\
K-Means (K=0.5$\sqrt{n}$)               & 4.73  &   84.50    & 25$x$\\
K-Means (K=$\sqrt{n}$)                  & 4.33  & 88.50  & 12$x$\\
K-Means (K=2$\sqrt{n}$)                 & 3.13  & 93.14 &   6$x$\\
K-Means (K=4$\sqrt{n}$)                 & 2.7   & 96.76 &   3.5$x$\\ 
Exhaustive list of $n$ &  0.0 & 100.0 &   $x$\\\bottomrule
\end{tabular}
}
\caption{Clustering results for author embeddings ($n$ is the number of papers per author, and $t$ is the target relevance count).  We report root mean squared error (RMSE) and how often the method is within approximately a factor of two of the true count (see text).  The summary embedding approaches (using K-means, for different values of K) provide large speedups at a small cost to accuracy.}
\label{tab:author_embed}
\end{table}

\section{User Study}
\label{sec:study}

We conducted a within-subjects user study over Zoom to evaluate the efficacy of~\sys in supporting exploratory search for the scientific literature domain. As a baseline, we compared to an existing production scientific literature navigation tool,~\semantic. As described in the previous sections, we used~\control as the basis for our system. It includes features for semantic and faceted search, viewing summaries / overviews of entity types (e.g., paper details page, author homepages, conference papers lists), and the ability to create and maintain personalized research feeds (paper recommenders), which we extended to function as lenses.

\subsection{Study Protocol}
\subsubsection{Task and Setup}
We asked our participants to conduct two short literature review tasks, one each using the features available via~\sys and~\control. To ensure a clear separation between the two conditions, we required that participants have curated at least two research feeds, one for each condition. Additionally, to avoid ordering effects, we randomized the order in which the two conditions were presented to each participant such that they were counterbalanced overall. The task entailed finding five new papers to add to the research feed and five new authors to follow based on relevance of their work to the research feed in question. After completing the task for each condition, participants reported their cognitive load using the NASA-TLX questionnaire~\cite{hart1988development} and evaluated system usability via the System Usability Scale~\cite{brooke1996sus}.

While participants' interactions with~\sys were limited to using one lens,~\sys features can be applied to a multi-lens setting. Once both conditions we completed, we asked participants to add their control condition research feed as a second lens to~\sys and reviewed the various features in this multi-lens setting. After giving them some time to explore the system in this new setting, we asked for their preference for using~\sys with one vs. multiple lenses at a time and any use-cases they felt were unique to single vs. multiple lenses. Participants also reported the usefulness of each new~\sys feature on a Likert scale of 0--5 (where 0 = not useful at all and 5 = extremely useful).

\subsubsection{Post-study Interviews}%
A subset of participants answered post-study questions about their preference for~\sys or~\control for completing the task---they were not required to answer this question with a clear preference. We had them explain their reasoning for the preferred system using individual features to compare between the two. They also provided design feedback on the visualizations. We followed up with open-ended questions about any changes they would make to~\sys features and if they could envision using~\sys in their usual literature review pipeline.

\begin{table*}[t]
\resizebox{0.8\textwidth}{!}{
\begin{tabular}{c|c|c|c|c|c|c|c|c|}
    \cline{2-9}
    & \multicolumn{2}{c|}{\begin{tabular}[c]{@{}c@{}}\textbf{Author Homepage} \\ \textbf{Views} \end{tabular}}
    & \multicolumn{2}{c|}{\textbf{Paper Views}} 
    & \multicolumn{2}{c|}{\begin{tabular}[c]{@{}c@{}}\textbf{Search Results} \\ \textbf{Page Views} \end{tabular}} 
    & \multicolumn{2}{c|}{\begin{tabular}[c]{@{}c@{}}\textbf{System-wide} \\ \textbf{Interactions} \end{tabular}} 
    \\ \cline{2-9} 
    \multicolumn{1}{c|}{} & \textbf{\sys} & \begin{tabular}[c]{@{}c@{}}\textbf{\textsc{Semantic} }\\\textbf{\textsc{Scholar}}\end{tabular} & \textbf{\sys} & \begin{tabular}[c]{@{}c@{}}\textbf{\textsc{Semantic}}\\\textbf{\textsc{Scholar}}\end{tabular} & \textbf{\sys} & \begin{tabular}[c]{@{}c@{}}\textbf{\textsc{Semantic}}\\\textbf{\textsc{Scholar}}\end{tabular} & \textbf{\sys} & \begin{tabular}[c]{@{}c@{}}\textbf{\textsc{Semantic}}\\\textbf{\textsc{Scholar}}\end{tabular} \\ \hline
    \multicolumn{1}{|c|}{\textbf{P1}}       & 10    & 10    & 5    & 1    & 10   & 8   & 232   & 116 \\
    \multicolumn{1}{|c|}{\textbf{P2}}      & 12    & 1     & 9    & 0    & 25   & 8   & 550   & 80  \\
    \multicolumn{1}{|c|}{\textbf{P3}}      & 32    & 5     & 0    & 7    & 12   & 18  & 187   & 255 \\
    \multicolumn{1}{|c|}{\textbf{P4}}      & 19    & 19    & 13   & 14   & 4    & 3   & 379   & 181 \\
    \multicolumn{1}{|c|}{\textbf{P5}}       & 12    & 16    & 0    & 0    & 6    & 6   & 274   & 152 \\
    \multicolumn{1}{|c|}{\textbf{P6}}       & 15    & 0    & 8    & 3    & 9   & 1   & 422   & 33 \\
    \multicolumn{1}{|c|}{\textbf{P7}}      & 16    & 1     & 0    & 0    & 37   & 2   & 550   & 5  \\
    \multicolumn{1}{|c|}{\textbf{P8}}      & 6    & 0     & 5    & 0    & 9   & 6  & 342   & 86 \\
    \multicolumn{1}{|c|}{\textbf{P9}}      & 35    & 3    & 1   & 4   & 11    & 2   & 433   & 27 \\
    \multicolumn{1}{|c|}{\textbf{P10}}       & 12    & 0    & 14  & 0    & 8    & 1   & 285   & 30 \\ \hline
    \multicolumn{1}{|c|}{\textbf{Mean}}  & \textbf{16.9 $\pm$ 9.5**}   & 5.5 $\pm$ 7.1     & \textbf{5.5 $\pm$ 5.4}   & 2.9 $\pm$ 4.6   & \textbf{13.1 $\pm$ 10.1*}    & 5.5 $\pm$ 5.2    & \textbf{365.4 $\pm$ 125.2***}  & 96.5 $\pm$ 80.1 \\\hline
\end{tabular}
}
\caption{Interaction logs for~\sys and~\semantic for the duration of the user study, including navigation to author homepages, papers, and search results page, and system-wide interactions (clicks and hovers).~\sys has higher values compared to~\semantic for all interaction logs. There were significant differences between~\sys and~\semantic for these numbers, indicated as: $*=p<.05$, $**=p<.01$, $***=p<.001$.}
\label{tab:site_logs}
\end{table*}

\subsection{Evaluation Methodology}
\textbf{Quantitative. }We compare ~\sys and~\control using three sets of metrics: high-level usability, and objective and subjective evaluation of papers and authors selected for the task. First, we capture \textit{high-level perceptions} of the task and the systems by measuring: (a) cognitive load using the NASA-TLX questionnaire~\cite{hart1988development}, (b) usability via the System Usability Scale, (c) user engagement based on log data and time spent using the system, and (d) perceived usefulness of individual~\sys features rated on a Likert scale of 0--5. Second, we analyze the papers and authors selected for the task using \textit{objective} metrics proposed by prior work on exploratory search and recommender systems. These include:
\begin{itemize}
    \item \textit{Relevance}, calculated as (1) the mean relevance score assigned by the recommender to the papers selected for the task, and (2) the mean number and proportion of relevant papers across the authors selected for the task. 
    \item \textit{Diversity}, calculated as a generalist-specialist score as defined by Anderson et al.~\cite{anderson2020algorithmic}. The generalist-specialist score calculates the mean distance between the centroid of a set of items and the individual items. Scores range from [-1--1] where a score of 1 represents a specialized set and -1 represents a general (i.e., diverse) set of items. 
    \item \textit{Novelty}, calculated as the mean distance between the paper and author lists selected for the task, and the centroid of the curated list of papers from the user's research feed. These values range from [0--1] where a higher score represents higher novelty. 
    \item \textit{Obviousness}, calculated as the mean number of citations for the papers and authors selected for the task. 
\end{itemize}

The diversity and novelty scores described above are calculated using SPECTER and tf-idf embeddings to represent a paper, and the mean SPECTER and tf-idf embeddings across all relevant papers to represent an author. We present the mean scores across the two types of embeddings. Finally, given the \textit{subjective} nature of literature review tasks, we verify relevance and novelty of papers and authors selected for the task by asking participants to rate these on a Likert scale of 0--5. All quantitative metrics were compared using paired t-tests; we include p-values for significant results.

\textbf{Qualitative. }We conducted an inductive thematic analysis on the auto-generated transcripts from the user study using the approach described by Corbin and Strauss~\cite{corbin1990grounded}. We first generated open codes for all dialogue instances, then organized them into axial codes using affinity diagramming.

\subsection{Participants and Data}~\label{sec:participants}
We recruited participants via email and word-of-mouth advertisement. All participants ($n=15$) were graduate students in computing-related fields. 4 were active~\control users and maintained 2--3 research feeds which they reviewed at least once a week. The remaining 11 participants were provided instructions on how to create research feeds, which they did at least one week before the study to ensure familiarity with their feeds.

We collected several types of data from all participants in the study, including lists of papers and authors selected for the literature review task, self-reports (e.g., cognitive load,~\sys feature usefulness, subjective evaluation of relevance and novelty, etc.), and log data. We report log data from ten participants; the remaining participants had specialized privacy settings that disallowed data recording.
Additionally, we asked a subset of ten participants to provide qualitative data in the form of post-study interviews with audio recorded for the duration of the study.

On average, the study took $\sim$34 mins with the base protocol and participants were compensated with \$12.5 for their time. Participants who additionally answered the qualitative questions took $\sim$48 mins and were compensated with \$25.

\subsection{Quantitative Results}
Our results are strongly positive.
First,~\sys\ users viewed and selected more content that was relevant according to the recommender, indicating engagement with \sys\ features that were intended to help users prioritize and adopt this content.
Using~\sys, participants \textit{selected} authors with a higher number of relevant papers (13$\pm$8 vs. 6$\pm$9 for~\control)\footnote{We report means and standard deviations in parentheses.} and a greater proportion of relevant papers (24.5$\pm$10.5\% for~\sys, 10.5$\pm$12\% for~\control), for the task. Similarly, the authors \textit{viewed} by participants (by visiting the author's homepage and, with~\sys, accessing the relevance overview chart by hovering over the author name) over the course of the task had a higher number of relevant papers (9.7$\pm$7 for~\sys vs. 3$\pm$1.7 for~\control); the proportions were substantially higher as well (21.8$\pm$9.8\% for~\sys, 5.5$\pm$4\% for~\control; $p<0.05$).

Furthermore, access to~\sys features made the literature review task less cognitively demanding and, in turn, promoted user engagement with the system. System-wide engagement with~\sys was \textit{4x} greater than~\control (Table~\ref{tab:site_logs}). People's self-reported cognitive load values were lower for~\sys\ (\sys: 2.4$\pm$1,~\control: 3.3$\pm$1.3; range 1--5). That these cognitive load values were lower for~\sys despite a strictly more complex interface indicates that our features were streamlining author-centric exploration as intended. Table~\ref{tab:site_logs} shows system-wide user engagement in the form of log data (e.g., clicks, pagination, navigating to different types of content lists). Coupled with the lower cognitive effort values and higher user engagement with the system, we considered more time spent on the task when using~\sys to be a positive (\sys: 15.8 mins,~\control: 10.76 mins, on average)---this was confirmed by participants in the post-study interview. 

\begin{figure}[t]
    \centering
    \includegraphics[width=\columnwidth]{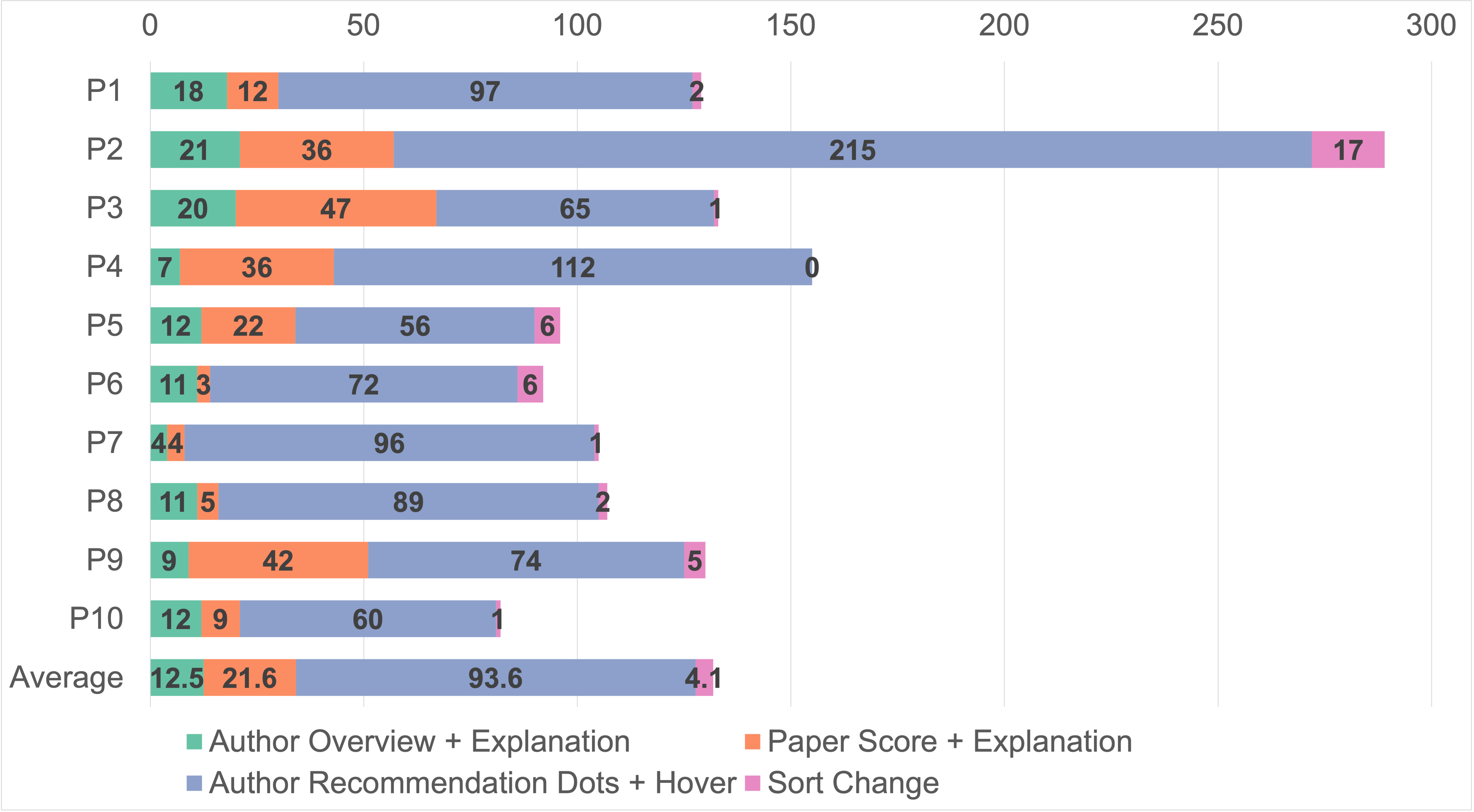}
    \caption{Usage counts for~\sys features show that the author recommendation dots were the most used feature, followed by paper relevance scores and explanations, author overview and explanation, and the sort functionality. For the first three, higher numbers are better since they indicate more feature use. For sort change, a lower number can be interpreted as better since it indicates that people liked the default sort order (based on a lens) and did not feel the need to change it to other options (e.g., recency, number of citations).}
    \label{fig:feature_counts}
\end{figure}

\begin{figure}[t]
    \centering
    \includegraphics[width=\columnwidth]{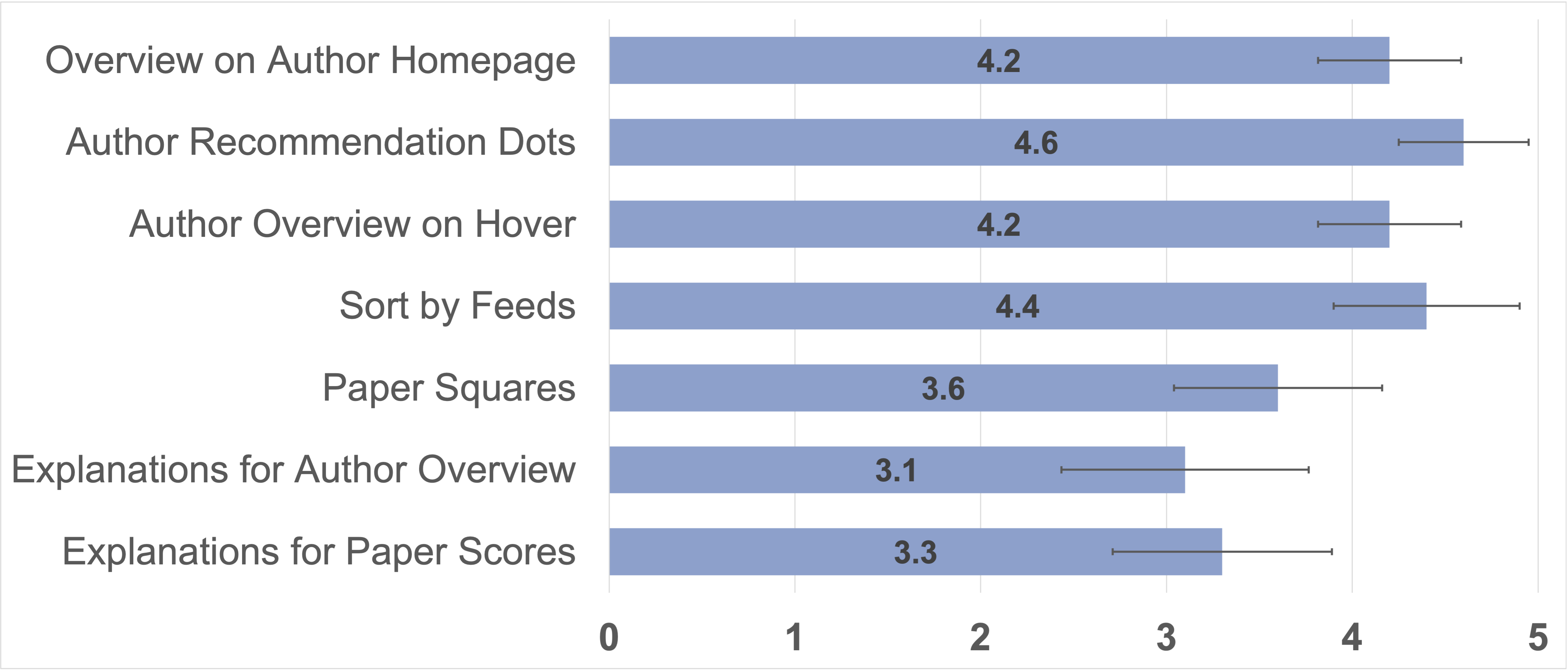}
    \caption{Usefulness of each~\sys feature, rated on a scale of 0--5. Author recommendation dots were rated as most useful, followed by author overview on hover, author overview on their homepage, and sort functionality, all of which received a rating of $>$4 out of 5. Bars show the mean value and horizontal lines show standard deviation.} %
    \label{fig:feature_usefulness}
\end{figure}

Looking at the author-related features in particular, 
we observed that people's interaction with authors (e.g., clicks, hovers, homepage views) was \textit{20x} greater when using \sys\ 
than with
\control\ (\sys: 123$\pm$47.4, \control: 5.5$\pm$7.1, $p<<0.001$). \sys promoted quick perusal and easy exploration of authors via author recommendation dots that drew attention to more relevant authors and the overview charts that could be easily accessed by hovering over the recommended authors (more details in qualitative results). Figure~\ref{fig:feature_counts} shows the usage counts for all~\sys features, corroborating the high usage counts for the author recommendation dots and hover feature. Even when considering only the author homepage views---a feature consistent across both conditions---user engagement numbers were \textit{3x} greater for~\sys ($p<0.01$; Table~\ref{tab:site_logs}-column 1). Given these results, it came as no surprise that these features were rated highly in our survey on the usefulness of~\sys features. Figure~\ref{fig:feature_usefulness} corroborates this with the nearly perfect rating for usefulness of author recommendation dots, followed by the rating for author overview on hover, overview on author homepage, and sort by feeds (all four features have average ratings of $>$4 out of 5).~\sys also received higher overall usability scores measured using the System Usability Scale (\sys: 84$\pm$8.6, ~\control: 77.3$\pm$12.5; range 0--100; $p<0.01$). 

There were only marginal differences between~\sys and \control for subjective and objective metrics calculated using the final list of papers selected for the task and similarly the final list of authors selected (except the objective relevance reported above); therefore, we omit these results here for brevity.\footnote{Comparison results for all of our metrics are included in the supplementary material.}

\subsection{Qualitative Themes}
These themes describe the qualitative data from the ten people who answered the additional post-study interview questions.

\sys promoted exploration by connecting multiple types of entities in the scientific literature KG using polymorphic lenses. Participants noted a preference for using~\sys over~\control for the study task (8 out of 10 participants preferred~\sys, 1 preferred~\control, 1 had no preference). They described the author-centric exploration enabled by polymorphic lenses as a ``new signal for finding relevant papers'' (P4), ``a trail of breadcrumbs to click through to find people who I should be looking at'' (P13), and ``a springboard for doing lit reviews'' (P3), among other things. Several of the benefits in the high-level usability results from the quantitative section were also consistent with the participants' perceived benefits of~\sys over~\control. Moreover, these benefits were tied to the design desiderata from the pilot surveys which we incorporated in our system design: support quick perusal of information (D1), embed new information in existing design (D2), and support building trust in the relevance information (D3). 

\subsubsection{Perceived Benefits of~\sys over~\control} \hfill \break
\indent \textbf{Lower cognitive effort.}~\sys features assisted in determining paper and author relevance, which in turn made the literature review task less onerous and required considerably less cognitive effort. All participants noted author recommendation dots as particularly helpful for this streamlined exploration: they termed these recommendation dots as ``hints'' (P2) or ``a narrowed path'' (P3) that made the task less cognitively demanding. The~\control condition was perceived as far more demanding since people had to either apply their own background knowledge or examine papers and authors in more detail to make the same relevance assessment. 
\begin{quote}
\textit{``I could have used a feature like that [author recommendation dots] right now [in the control condition]. Now this is getting taxing and I don't like it. The first one [\sys condition] was much easier. So this [control condition] is kind of demanding and I'm frustrated and I'm thinking I'm wasting my time without those features to streamline the task.'' (P5)} 
\end{quote}

\textbf{Higher user engagement.}~\sys served its intended purpose in promoting more exploration across the search system. One concrete example of this was the author recommendation dots and the overview charts presented upon hovering over them, which prompted more interest in authors. Participants viewed more author homepages and in turn found more relevant papers with ease: 
\begin{quote}
    \textit{`I'm drawn towards these two [recommended authors].... Let's see what they have. Wow twenty five papers [by this author] are related to my feed because VR, avatars, immersive [as explanations]. I can't believe I've never heard of this author before.''} (P11)
\end{quote}
More generally, several participants commented on~\sys features being more fun and exciting to use, which made them want to continue exploring content (P4) whereas the~\control condition was challenging due to a lack of similar features (P14). Even our participant with no clear preference for using~\sys vs.~\control felt that the former was more exciting because ``it [the new features] definitely made it more fun %
\ldots and engaging.'' (P11).

\textbf{Nuanced author selection.} \sys features and visualizations presented author information such that people could conduct a more nuanced evaluation of relevance, enabling them to select authors that might otherwise be overlooked. Recall that the author recommendation dots were designed to accommodate new authors in a domain to avoid rich-get-richer effects (Section~\ref{author_recs}). Using~\sys, participants were able to find authors that were new to a research domain or junior e.g., graduate students:
\begin{quote}
    \textit{``The cool part was that I found one of the authors who just had two papers, which I thought was interesting. Hopefully if the way it [\textsc{FeedLens}] recommends authors it doesn't necessarily just do it based on [total] number of papers alone. That's a good thing because obviously there are issues with just [total] number of papers if you end up not finding people who are just starting their research. I know this person only has two papers but when it’s relevant, it’s relevant.''} (P1)
\end{quote}

\subsubsection{Underlying Factors Supporting~\sys's Performance}\hfill \break
\indent\textbf{Quick evaluation of relevance.}~\sys features were optimized for quick evaluation of relevance for individual papers and authors: the relevance-based ranking of papers provided by our default sort option resulted in participants anticipating that, if an author was relevant, they would see a large number of squares on the first page of results on the author homepage, since these results were sorted by relevance to their lens by default. This did not necessarily require internalizing the relevance score for each paper, making it an efficient perusal in terms of time and cognitive effort. This sort order was particularly useful for finding relevant papers by prolific authors who may have contributed to multiple research domains. For example, a professor ``who has like one or two students that do research [on my topic], so it's hard to usually tell relevance, but with the sorted squares there's an easy signal'' (P4). With no equivalent way to get this relevance information in~\control, participants spent (what they perceived as) significantly more time evaluating author relevance in that case.

\textbf{Ease of comparing entities. }
\sys features included visual elements (e.g., green circles and squares) to indicate relevance along with numerical values (e.g., relevant paper counts, relevance scores) to enable precise comparisons. Participants used these features for comparisons in both intended and unintended ways. Access to the author overview chart by hovering over a recommended author dot was intended to enable efficient high-level comparison between authors, and was used as such. This efficiency was appreciated by our participants---as noted in our quantitative results, author-centric engagement via~\sys is 20x the number for~\control: 
\begin{quote}
    \textit{``The sheer quantity of information I can easily parse from this one page... so useful. You'd think that looking at [all] authors on a page [vs. papers] would be daunting but I can be selective with a simple first pass over all the overviews. Certainly makes it appealing to look at authors, which I don't do as often as I should.''} (P13)
\end{quote}

One unintended benefit of the author recommendation dots feature was in quickly assessing which \textit{papers} were more relevant. P2 explained that ``the nice thing about seeing these little green dots is that I can more quickly figure out which of these papers to look at. Since this paper doesn't have any green dots, maybe it's probably not relevant. So I think I'm not going to look at it as much.'' Several participants (4 out of 10) explicitly noted following a similar approach and were frustrated by not being able to do the same with~\control: ``Oh my. My first thought is that I can't easily find if the paper is relevant without the dots.'' (P5).

\textbf{Streamlined browsing. }All participants commented on the streamlined nature of exploration supported by~\sys which made the search task engaging and fun instead of stressful because of information overload. With our progressive summarization approach, author recommendation dots were primarily responsible for this streamlined browsing experience. All lists of papers across the search interface were pervasively customized to highlight the relevant authors using these dots. As a result, it was easy to quickly skim the collections of papers and authors on a page: ``This is actually pretty cool. This is way more useful [than~\textsc{S2}], like I know stuff immediately like this one [author] seems very relevant. With just a glance.'' (P14). One participant also highlighted the use of paper squares for streamlined browsing, a use case that we had not considered when designing this feature:
\begin{quote}
    \textit{``I'm going to just sort by recency to see what they've done recently… you know what, now I really like these squares. I can immediately tell which of this author’s recent papers are relevant to me.''} (P1)
\end{quote}

\begin{figure*}[t]
    \centering
    \includegraphics[width=0.85\textwidth]{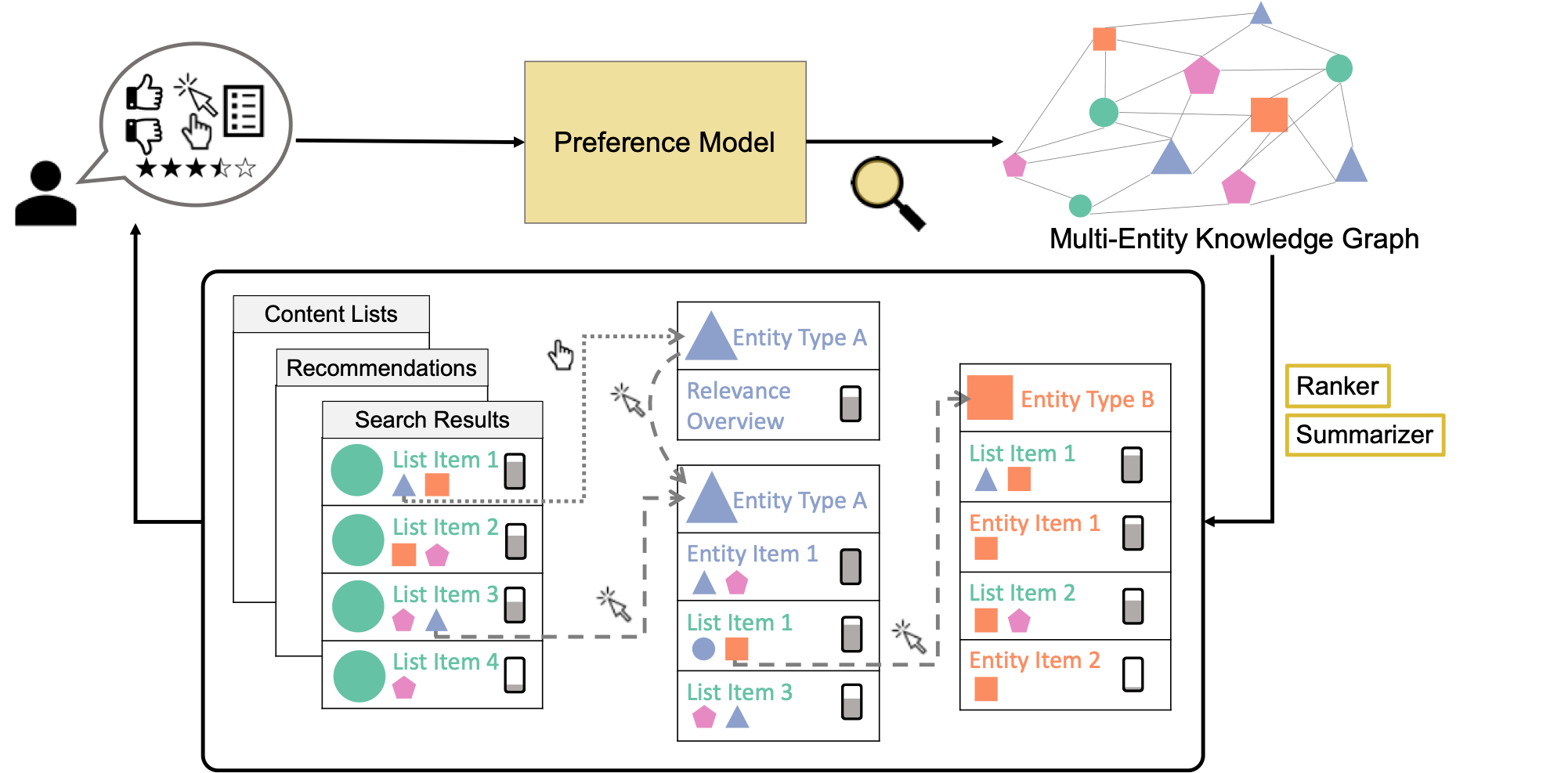}
    \caption{
    Polymorphic lenses, our technique for improving exploratory search over knowledge graphs, capitalize on existing preference models used in search and recommender systems, and generalize their application from a base entity (e.g., papers in the scientific literature KG) to \textit{all} types of entities in a knowledge graph (e.g., authors, conferences, institutions). With this simple generalization, all entities and collections of entities can be ranked and summarized based on their relevance to the lens. This enables new exploration opportunities: content lists can be viewed sorted by relevance to a lens, and all types of entities can be interconnected due to this shared notion of relevance. Here, we see a search results list of base entities (green circles) ranked by relevance scores (gray fill indicators); a summary of Entity Type A's relevance can be previewed on hovering over a triangle (hover action shown as a dotted line from the triangle attribute of the first base entity), and the full list of items associated with Entity A (below the hover modal) can be similarly ranked, summarized, and interconnected (e.g., by clicking through to the list of items associated with Entity B on the right).
    }
    \label{fig:overview}
\end{figure*}

\subsubsection{Feedback on~\sys Design}
Most participants felt that the explanations, while helpful in some ways, were generally too high-level. They did not provide additional information that the participant could not have gleaned from reading a paper's title and abstract. One participant noted this as the advantage of explanations---they provided immediate context without needing to read more about a paper. Overall, while explanations were perceived as less useful than the other features, participants did not have a clear idea for what they would have liked to see instead of the high-level keywords provided. Explanations were important as a means for intelligibility and establishing trust in the relevance scores shown for each paper, but more design and technical work is necessary before they can be used as intended.  

Participants also caught on to a conflict between our design considerations for the visualizations designed for paper overview. They wanted a visual that allowed quick perusal (e.g., the squares in the paper row, in our design) but also more intelligibility (e.g., more information that simple explanations), without cluttering the already dense paper row. They did not have feedback on how to accomplish this, but we hope to explore this further in future work. For polymorphic lenses to be beneficial in existing systems, it is imperative that we find additional ways of seamlessly merging these new exploratory endpoints in existing interface designs.

\subsubsection{Subjective Notions of Task Success}\label{sec:task}
We know of no objective metric that indicates a ``successful'' exploratory search. Instead, success is continually re-defined because exploratory search relies on knowledge acquisition and synthesis, both of which are dynamic processes that, in turn, can update the original search query~\cite{noy2019industry,white2006supporting}. As such, exploratory search, even for a shortened task like ours, requires considerable cognitive effort.

The cognitive effort required for evaluating outputs of exploratory search was easily observable when participants struggled with defining relevance and novelty---our subjective metrics---for rating the 5 papers and authors selected for the literature review task. Everyone made assumptions in defining these success metrics, and these were inconsistent across participants. For example, while some participants evaluated author relevance based on the number of relevant papers from that author, others also considered whether the author published in good quality journals---quality, again, being subjective. One participant even considered two authors to be similarly relevant if ``[one] author wrote [even just] the one paper that inspired my thesis or if someone wrote 10 papers on my feed topic.'' (P6). Another relied on signals such as ``having met the authors being considered in person or have already read multiple papers by them'' (P5). This evaluation of relevance and other metrics is implicit in all exploratory search tasks, given that people select papers and authors by defining (and iteratively re-defining) relevance for their search query.
\section{Discussion}
\textbf{Synthesis of results. }
In sum, our results show that people preferred using~\sys primarily because its features allowed for more exploration opportunities when conducting a literature review.~\sys promoted relevant authors as new exploratory, navigational end-points, which resulted in the selection of more relevant authors for the task.~\sys features were optimized for quick perusal of information and, as intended, required less cognitive effort for completing the task.~\sys also encouraged more user engagement and, in turn, more time spent on exploring the system. The final list of papers and authors selected for the task showed only marginal differences on other commonly-used objective (e.g., diversity, novelty) and subjective (e.g., user-reported relevance and novelty) metrics. Perhaps the clearest signal of~\sys's usability and the benefits afforded by its design is that several people wanted to be able to use it immediately for their literature review pipelines:
\begin{quote}
    \textit{``I really like the features, and that's because for one of the feeds [with \textsc{FeedLens}], I was able to quickly find a lot of papers for which I felt, `oops, I should have read these papers.' And if I am able to do that in 15 minutes of my time, that's extremely useful. So, if anything, that makes me want to use it more. It's pretty cool. I hope it gets launched soon because I want to use it.''} (P5)
\end{quote}

\textbf{Application to new domains. }
While \sys\ is a system for improved exploratory search over a scientific literature knowledge graph, the idea of polymorphic lenses is general, applying to any domain represented as a KG and using a preference model (Figure~\ref{fig:overview}). 
 For example, one could apply our technique to Yelp\footnote{\url{https://www.yelp.com/}} and offer new exploration end-points such as cities summarized and ranked based on a user's preferred cuisines or hobbies, or restaurants summarized based on customer reviews from the top-k similar users. Similarly, applied to the preference models captured by online streaming and content recommendation platforms like Netflix,\footnote{\url{https://www.netflix.com}} polymorphic lenses could rank actors and directors by the users' favorite genres, or even other users based on shared ratings of movies. We hope that the simplicity and generality of the polymorphic lenses approach will allow for widespread application, and enable novel and interesting exploration endpoints for various types and collections of entities across the many KGs commonly used online.

\section{Limitations}
While we have described possible extensions of our polymorphic lenses technique to other domains, we test it in one domain. It could be that the average person conducting a literature review spends more time exploring a KG; the benefits of our technique might not be as remarkable in other domains. Our numbers are also limited to a sample size of 15 and, for some metrics, 10 due to the privacy settings of the participants' browsers. These trends might look different with larger sample sizes. Additionally, our evaluation numbers are dependent on how well people maintain their feeds (i.e., the lenses). While we were assured of this for the purposes of the study, un-curated or ill-maintained feeds from other~\control users might not be as helpful. Our user study also validated the approach using results for one higher-level entity type, authors. We limited ourselves to one domain, higher-level entity type, and production system, \control, in part due to the development time and resources that are needed to add features to an existing system. 

Though our user study provides a nuanced comparison between \sys and~\control, a within-subjects design cannot ensure separation between conditions, especially in a case where one system (\sys) extends the features of another (\control). We tried to combat these issues by using different research feeds for each condition and randomizing the order in which the conditions were presented, but we cannot speak to the full generalizability of our results without controlled experiments with larger sample sizes. However, greater levels of control would require research feeds that are consistent across participants, which is infeasible in a personalized setting like ours. Future work must consider the tradeoff between the internal and external validity advantages that could be achieved by keeping lenses consistent across participants (e.g., by defining them as a part of the experiment), and the ecological validity concerns that would be introduced by not using people's personalized feeds as lenses. Similar ecological validity challenges also exist due to the shortened nature of our literature review task. Our observations suggest that people had to engage in the usual sensemaking and synthesizing needed for an exploratory search task, despite our task being shortened for feasibility (Section~\ref{sec:task}). However, testing this in the wild might lead to different outcomes.

Finally, \sys provides support to users through relevance indicators that help to focus user attention in productive ways. While this approach offers benefits such as reduced cognitive load, a concern with any such approach is that redirecting attention patterns may produce unintended harm. Possible harms in the scientific literature domain include harm to the users who may skip less relevant papers that are nonetheless beneficial (e.g., provide scientific inspiration), or harm to the authors of those papers, who may not receive a citation they might have otherwise received. We have taken several precautions to mitigate these potential harms. \sys\ does not hide papers or authors; instead, it provides additional information to assist users. Further, we introduce a mechanism to balance recommending some (randomly selected) authors with a small number of relevant publications without overly cluttering the interface (which could potentially increase cognitive load). Indeed, we find that our participants explore more (relevant) authors without introducing bias towards well known authors, as measured by no significant difference between the author obviousness metric for~\sys and~\control during user study. We believe that this is a net benefit for users and authors. However, further study is needed to fully characterize any effects on the distribution of papers and authors visited by the user.

\section{Conclusion}

Many content-based applications explicitly recommend new content to users. Typically, these recommendations stem from a personalized profile, which implements a preference model over a base type in an underlying knowledge graph. For example, Netflix records which movies a user liked and scientific literature applications know which papers a researcher saves to their library. Previous work has demonstrated that a ``lens'' metaphor can enable ad-hoc preference models to improve exploratory search in the domain of Web search~\cite{chang2019searchlens}. We extend that idea to the notion of polymorphic lenses, showing how one may use the linkage structure of the knowledge graph to create preference scores for all types of entities. Polymorphic lenses allow one to personalize entity lists all across the application and to summarize an entity in terms of key elements from its neighborhood in the graph. Further, compared to the ad-hoc approach, our approach reduces startup cost by re-using existing preference models already present in such applications.

To demonstrate the advantages of our approach, we implemented polymorphic lenses in an existing production system which relies on a billion-node literature graph of authors, papers, venues, and institutions. Results from a within-subjects user study comparing our system (\sys) to the existing production system (\semantic) show
that \sys allowed people to explore more content in the same time with lower cognitive load. Furthermore, users gave \sys\ a higher usability score. We conclude that polymorphic lenses expand the benefits of personalization, improving sensemaking, filtering, and organizing aspects of exploratory search compared to existing literature review practices. They are a simple and general approach that likely extends to applications over diverse knowledge graphs in many domains. 

\begin{acks}
We are  grateful to Alex Schokking, Chris Wilhelm, Paul Sayre, Matt Latzke, and the entire Semantic Scholar team for their help with designing \&\ integrating our features into a production system. We also thank Tongshuang Wu,  Mitchell Gordon, Joseph Chang, and our anonymous reviewers for very helpful feedback and suggestions.
This  work was supported, in part, by  NSF Grant OIA-2033558, NSF RAPID grant 2040196,  and ONR grant {N00014-18-1-2193}.
\end{acks}

\bibliographystyle{ACM-Reference-Format}
\bibliography{0x_references}

%%% -*-BibTeX-*-
%%% Do NOT edit. File created by BibTeX with style
%%% ACM-Reference-Format-Journals [18-Jan-2012].

\begin{thebibliography}{51}

%%% ====================================================================
%%% NOTE TO THE USER: you can override these defaults by providing
%%% customized versions of any of these macros before the \bibliography
%%% command.  Each of them MUST provide its own final punctuation,
%%% except for \shownote{}, \showDOI{}, and \showURL{}.  The latter two
%%% do not use final punctuation, in order to avoid confusing it with
%%% the Web address.
%%%
%%% To suppress output of a particular field, define its macro to expand
%%% to an empty string, or better, \unskip, like this:
%%%
%%% \newcommand{\showDOI}[1]{\unskip}   % LaTeX syntax
%%%
%%% \def \showDOI #1{\unskip}           % plain TeX syntax
%%%
%%% ====================================================================

\ifx \showCODEN    \undefined \def \showCODEN     #1{\unskip}     \fi
\ifx \showDOI      \undefined \def \showDOI       #1{#1}\fi
\ifx \showISBNx    \undefined \def \showISBNx     #1{\unskip}     \fi
\ifx \showISBNxiii \undefined \def \showISBNxiii  #1{\unskip}     \fi
\ifx \showISSN     \undefined \def \showISSN      #1{\unskip}     \fi
\ifx \showLCCN     \undefined \def \showLCCN      #1{\unskip}     \fi
\ifx \shownote     \undefined \def \shownote      #1{#1}          \fi
\ifx \showarticletitle \undefined \def \showarticletitle #1{#1}   \fi
\ifx \showURL      \undefined \def \showURL       {\relax}        \fi
% The following commands are used for tagged output and should be
% invisible to TeX
\providecommand\bibfield[2]{#2}
\providecommand\bibinfo[2]{#2}
\providecommand\natexlab[1]{#1}
\providecommand\showeprint[2][]{arXiv:#2}

\bibitem[Anderson et~al\mbox{.}(2020)]%
        {anderson2020algorithmic}
\bibfield{author}{\bibinfo{person}{Ashton Anderson}, \bibinfo{person}{Lucas
  Maystre}, \bibinfo{person}{Ian Anderson}, \bibinfo{person}{Rishabh Mehrotra},
  {and} \bibinfo{person}{Mounia Lalmas}.} \bibinfo{year}{2020}\natexlab{}.
\newblock \showarticletitle{Algorithmic effects on the diversity of consumption
  on spotify}. In \bibinfo{booktitle}{\emph{Proceedings of The Web Conference
  2020}}. \bibinfo{pages}{2155--2165}.
\newblock


\bibitem[Baldonado and Winograd(1997)]%
        {baldonado1997sensemaker}
\bibfield{author}{\bibinfo{person}{Michelle Q~Wang Baldonado} {and}
  \bibinfo{person}{Terry Winograd}.} \bibinfo{year}{1997}\natexlab{}.
\newblock \showarticletitle{SenseMaker: An information-exploration interface
  supporting the contextual evolution of a user's interests}. In
  \bibinfo{booktitle}{\emph{Proceedings of the ACM SIGCHI Conference on Human
  factors in computing systems}}. \bibinfo{pages}{11--18}.
\newblock


\bibitem[Beltagy et~al\mbox{.}(2019)]%
        {Beltagy2019SciBERT}
\bibfield{author}{\bibinfo{person}{Iz Beltagy}, \bibinfo{person}{Kyle Lo},
  {and} \bibinfo{person}{Arman Cohan}.} \bibinfo{year}{2019}\natexlab{}.
\newblock \showarticletitle{SciBERT: Pretrained Language Model for Scientific
  Text}. In \bibinfo{booktitle}{\emph{EMNLP}}.
\newblock
\showeprint{arXiv:1903.10676}


\bibitem[Bernard et~al\mbox{.}(2013)]%
        {bernard2013motionexplorer}
\bibfield{author}{\bibinfo{person}{J{\"u}rgen Bernard}, \bibinfo{person}{Nils
  Wilhelm}, \bibinfo{person}{Bj{\"o}rn Kr{\"u}ger}, \bibinfo{person}{Thorsten
  May}, \bibinfo{person}{Tobias Schreck}, {and} \bibinfo{person}{J{\"o}rn
  Kohlhammer}.} \bibinfo{year}{2013}\natexlab{}.
\newblock \showarticletitle{Motionexplorer: Exploratory search in human motion
  capture data based on hierarchical aggregation}.
\newblock \bibinfo{journal}{\emph{IEEE transactions on visualization and
  computer graphics}} \bibinfo{volume}{19}, \bibinfo{number}{12}
  (\bibinfo{year}{2013}), \bibinfo{pages}{2257--2266}.
\newblock


\bibitem[Bollacker et~al\mbox{.}(2008)]%
        {Bollacker2008FreebaseAC}
\bibfield{author}{\bibinfo{person}{Kurt~D. Bollacker}, \bibinfo{person}{Colin
  Evans}, \bibinfo{person}{Praveen~K. Paritosh}, \bibinfo{person}{Tim Sturge},
  {and} \bibinfo{person}{Jamie Taylor}.} \bibinfo{year}{2008}\natexlab{}.
\newblock \showarticletitle{Freebase: a collaboratively created graph database
  for structuring human knowledge}. In \bibinfo{booktitle}{\emph{SIGMOD
  Conference}}.
\newblock


\bibitem[Brooke(1996)]%
        {brooke1996sus}
\bibfield{author}{\bibinfo{person}{John Brooke}.}
  \bibinfo{year}{1996}\natexlab{}.
\newblock \showarticletitle{Sus: a “quick and dirty’usability}.
\newblock \bibinfo{journal}{\emph{Usability evaluation in industry}}
  \bibinfo{volume}{189} (\bibinfo{year}{1996}).
\newblock


\bibitem[Buxton(2010)]%
        {buxton2010sketching}
\bibfield{author}{\bibinfo{person}{Bill Buxton}.}
  \bibinfo{year}{2010}\natexlab{}.
\newblock \bibinfo{booktitle}{\emph{Sketching user experiences: getting the
  design right and the right design}}.
\newblock \bibinfo{publisher}{Morgan kaufmann}.
\newblock


\bibitem[Cao et~al\mbox{.}(2019)]%
        {cao2019unifying}
\bibfield{author}{\bibinfo{person}{Yixin Cao}, \bibinfo{person}{Xiang Wang},
  \bibinfo{person}{Xiangnan He}, \bibinfo{person}{Zikun Hu}, {and}
  \bibinfo{person}{Tat-Seng Chua}.} \bibinfo{year}{2019}\natexlab{}.
\newblock \showarticletitle{Unifying knowledge graph learning and
  recommendation: Towards a better understanding of user preferences}. In
  \bibinfo{booktitle}{\emph{The world wide web conference}}.
  \bibinfo{pages}{151--161}.
\newblock


\bibitem[Chang et~al\mbox{.}(2019)]%
        {chang2019searchlens}
\bibfield{author}{\bibinfo{person}{Joseph~Chee Chang}, \bibinfo{person}{Nathan
  Hahn}, \bibinfo{person}{Adam Perer}, {and} \bibinfo{person}{Aniket Kittur}.}
  \bibinfo{year}{2019}\natexlab{}.
\newblock \showarticletitle{SearchLens: Composing and capturing complex user
  interests for exploratory search}. In \bibinfo{booktitle}{\emph{Proceedings
  of the 24th International Conference on Intelligent User Interfaces}}.
  \bibinfo{pages}{498--509}.
\newblock


\bibitem[Chellappa and Sin(2005)]%
        {chellappa2005personalization}
\bibfield{author}{\bibinfo{person}{Ramnath~K Chellappa} {and}
  \bibinfo{person}{Raymond~G Sin}.} \bibinfo{year}{2005}\natexlab{}.
\newblock \showarticletitle{Personalization versus privacy: An empirical
  examination of the online consumer’s dilemma}.
\newblock \bibinfo{journal}{\emph{Information technology and management}}
  \bibinfo{volume}{6}, \bibinfo{number}{2} (\bibinfo{year}{2005}),
  \bibinfo{pages}{181--202}.
\newblock


\bibitem[Cohan et~al\mbox{.}(2020)]%
        {specter2020cohan}
\bibfield{author}{\bibinfo{person}{Arman Cohan}, \bibinfo{person}{Sergey
  Feldman}, \bibinfo{person}{Iz Beltagy}, \bibinfo{person}{Doug Downey}, {and}
  \bibinfo{person}{Daniel~S. Weld}.} \bibinfo{year}{2020}\natexlab{}.
\newblock \showarticletitle{{SPECTER: Document-level Representation Learning
  using Citation-informed Transformers}}. In \bibinfo{booktitle}{\emph{ACL}}.
\newblock


\bibitem[Corbin and Strauss(1990)]%
        {corbin1990grounded}
\bibfield{author}{\bibinfo{person}{Juliet~M Corbin} {and}
  \bibinfo{person}{Anselm Strauss}.} \bibinfo{year}{1990}\natexlab{}.
\newblock \showarticletitle{Grounded theory research: Procedures, canons, and
  evaluative criteria}.
\newblock \bibinfo{journal}{\emph{Qualitative sociology}} \bibinfo{volume}{13},
  \bibinfo{number}{1} (\bibinfo{year}{1990}), \bibinfo{pages}{3--21}.
\newblock


\bibitem[Eslami et~al\mbox{.}(2015)]%
        {eslami2015always}
\bibfield{author}{\bibinfo{person}{Motahhare Eslami}, \bibinfo{person}{Aimee
  Rickman}, \bibinfo{person}{Kristen Vaccaro}, \bibinfo{person}{Amirhossein
  Aleyasen}, \bibinfo{person}{Andy Vuong}, \bibinfo{person}{Karrie Karahalios},
  \bibinfo{person}{Kevin Hamilton}, {and} \bibinfo{person}{Christian Sandvig}.}
  \bibinfo{year}{2015}\natexlab{}.
\newblock \showarticletitle{" I always assumed that I wasn't really that close
  to [her]" Reasoning about Invisible Algorithms in News Feeds}. In
  \bibinfo{booktitle}{\emph{Proceedings of the 33rd annual ACM conference on
  human factors in computing systems}}. \bibinfo{pages}{153--162}.
\newblock


\bibitem[Fishkin(2019)]%
        {fishkin2019less}
\bibfield{author}{\bibinfo{person}{Rand Fishkin}.}
  \bibinfo{year}{2019}\natexlab{}.
\newblock \showarticletitle{Less than half of google searches now result in a
  click}.
\newblock \bibinfo{journal}{\emph{Sparktoro Blog}} (\bibinfo{year}{2019}).
\newblock


\bibitem[Fogarty(2017)]%
        {fogarty2017code}
\bibfield{author}{\bibinfo{person}{James Fogarty}.}
  \bibinfo{year}{2017}\natexlab{}.
\newblock \showarticletitle{Code and contribution in interactive systems
  research}. In \bibinfo{booktitle}{\emph{Workshop HCITools: Strategies and
  Best Practices for Designing, Evaluating and Sharing Technical HCI Toolkits
  at CHI}}.
\newblock


\bibitem[Fox et~al\mbox{.}(2006)]%
        {fox2006exploring}
\bibfield{author}{\bibinfo{person}{Edward~A Fox}, \bibinfo{person}{Fernando~Das
  Neves}, \bibinfo{person}{Xiaoyan Yu}, \bibinfo{person}{Rao Shen},
  \bibinfo{person}{Seonho Kim}, {and} \bibinfo{person}{Weiguo Fan}.}
  \bibinfo{year}{2006}\natexlab{}.
\newblock \showarticletitle{Exploring the computing literature with
  visualization and stepping stones \& pathways}.
\newblock \bibinfo{journal}{\emph{Commun. ACM}} \bibinfo{volume}{49},
  \bibinfo{number}{4} (\bibinfo{year}{2006}), \bibinfo{pages}{52--58}.
\newblock


\bibitem[Gil(2022)]%
        {gil2022will}
\bibfield{author}{\bibinfo{person}{Yolanda Gil}.}
  \bibinfo{year}{2022}\natexlab{}.
\newblock \showarticletitle{Will AI write scientific papers in the future?}
\newblock \bibinfo{journal}{\emph{AI Magazine}} \bibinfo{volume}{42},
  \bibinfo{number}{4} (\bibinfo{year}{2022}), \bibinfo{pages}{3--15}.
\newblock


\bibitem[Glowacka et~al\mbox{.}(2013)]%
        {Glowacka2013DirectingES}
\bibfield{author}{\bibinfo{person}{D. Glowacka}, \bibinfo{person}{Tuukka
  Ruotsalo}, \bibinfo{person}{Ksenia Konyushkova}, \bibinfo{person}{K.
  Athukorala}, \bibinfo{person}{Samuel Kaski}, {and} \bibinfo{person}{G.
  Jacucci}.} \bibinfo{year}{2013}\natexlab{}.
\newblock \showarticletitle{Directing exploratory search: reinforcement
  learning from user interactions with keywords}. In
  \bibinfo{booktitle}{\emph{IUI '13}}.
\newblock


\bibitem[Grevet and Gilbert(2015)]%
        {grevet2015piggyback}
\bibfield{author}{\bibinfo{person}{Catherine Grevet} {and}
  \bibinfo{person}{Eric Gilbert}.} \bibinfo{year}{2015}\natexlab{}.
\newblock \showarticletitle{Piggyback prototyping: Using existing, large-scale
  social computing systems to prototype new ones}. In
  \bibinfo{booktitle}{\emph{Proceedings of the 33rd Annual ACM Conference on
  Human Factors in Computing Systems}}. \bibinfo{pages}{4047--4056}.
\newblock


\bibitem[Guo et~al\mbox{.}(2020)]%
        {guo2020survey}
\bibfield{author}{\bibinfo{person}{Qingyu Guo}, \bibinfo{person}{Fuzhen
  Zhuang}, \bibinfo{person}{Chuan Qin}, \bibinfo{person}{Hengshu Zhu},
  \bibinfo{person}{Xing Xie}, \bibinfo{person}{Hui Xiong}, {and}
  \bibinfo{person}{Qing He}.} \bibinfo{year}{2020}\natexlab{}.
\newblock \showarticletitle{A survey on knowledge graph-based recommender
  systems}.
\newblock \bibinfo{journal}{\emph{IEEE Transactions on Knowledge and Data
  Engineering}} (\bibinfo{year}{2020}).
\newblock


\bibitem[Hart and Staveland(1988)]%
        {hart1988development}
\bibfield{author}{\bibinfo{person}{Sandra~G Hart} {and}
  \bibinfo{person}{Lowell~E Staveland}.} \bibinfo{year}{1988}\natexlab{}.
\newblock \showarticletitle{Development of NASA-TLX (Task Load Index): Results
  of empirical and theoretical research}.
\newblock In \bibinfo{booktitle}{\emph{Advances in psychology}}.
  Vol.~\bibinfo{volume}{52}. \bibinfo{publisher}{Elsevier},
  \bibinfo{pages}{139--183}.
\newblock


\bibitem[Hearst(2006)]%
        {hearst2006clustering}
\bibfield{author}{\bibinfo{person}{Marti~A Hearst}.}
  \bibinfo{year}{2006}\natexlab{}.
\newblock \showarticletitle{Clustering versus faceted categories for
  information exploration}.
\newblock \bibinfo{journal}{\emph{Commun. ACM}} \bibinfo{volume}{49},
  \bibinfo{number}{4} (\bibinfo{year}{2006}), \bibinfo{pages}{59--61}.
\newblock


\bibitem[Hecht et~al\mbox{.}(2012)]%
        {hecht2012explanatory}
\bibfield{author}{\bibinfo{person}{Brent Hecht}, \bibinfo{person}{Samuel~H
  Carton}, \bibinfo{person}{Mahmood Quaderi}, \bibinfo{person}{Johannes
  Sch{\"o}ning}, \bibinfo{person}{Martin Raubal}, \bibinfo{person}{Darren
  Gergle}, {and} \bibinfo{person}{Doug Downey}.}
  \bibinfo{year}{2012}\natexlab{}.
\newblock \showarticletitle{Explanatory semantic relatedness and explicit
  spatialization for exploratory search}. In
  \bibinfo{booktitle}{\emph{Proceedings of the 35th international ACM SIGIR
  conference on Research and development in information retrieval}}.
  \bibinfo{pages}{415--424}.
\newblock


\bibitem[Heist et~al\mbox{.}(2020)]%
        {heist2020knowledge}
\bibfield{author}{\bibinfo{person}{Nicolas Heist}, \bibinfo{person}{Sven
  Hertling}, \bibinfo{person}{Daniel Ringler}, {and} \bibinfo{person}{Heiko
  Paulheim}.} \bibinfo{year}{2020}\natexlab{}.
\newblock \bibinfo{title}{Knowledge Graphs on the Web-An Overview.}
\newblock
\newblock


\bibitem[Jawaheer et~al\mbox{.}(2014)]%
        {jawaheer2014modeling}
\bibfield{author}{\bibinfo{person}{Gawesh Jawaheer}, \bibinfo{person}{Peter
  Weller}, {and} \bibinfo{person}{Patty Kostkova}.}
  \bibinfo{year}{2014}\natexlab{}.
\newblock \showarticletitle{Modeling user preferences in recommender systems: A
  classification framework for explicit and implicit user feedback}.
\newblock \bibinfo{journal}{\emph{ACM Transactions on Interactive Intelligent
  Systems (TiiS)}} \bibinfo{volume}{4}, \bibinfo{number}{2}
  (\bibinfo{year}{2014}), \bibinfo{pages}{1--26}.
\newblock


\bibitem[Kammerer et~al\mbox{.}(2009)]%
        {kammerer2009signpost}
\bibfield{author}{\bibinfo{person}{Yvonne Kammerer}, \bibinfo{person}{Rowan
  Nairn}, \bibinfo{person}{Peter Pirolli}, {and} \bibinfo{person}{Ed~H Chi}.}
  \bibinfo{year}{2009}\natexlab{}.
\newblock \showarticletitle{Signpost from the masses: learning effects in an
  exploratory social tag search browser}. In
  \bibinfo{booktitle}{\emph{Proceedings of the SIGCHI conference on human
  factors in computing systems}}. \bibinfo{pages}{625--634}.
\newblock


\bibitem[Lee et~al\mbox{.}(2020)]%
        {lee2020explanationbased}
\bibfield{author}{\bibinfo{person}{Benjamin Charles~Germain Lee},
  \bibinfo{person}{Kyle Lo}, \bibinfo{person}{Doug Downey}, {and}
  \bibinfo{person}{Daniel~S. Weld}.} \bibinfo{year}{2020}\natexlab{}.
\newblock \bibinfo{title}{Explanation-Based Tuning of Opaque Machine Learners
  with Application to Paper Recommendation}.
\newblock
\newblock
\showeprint[arxiv]{2003.04315}~[cs.IR]


\bibitem[Luciani et~al\mbox{.}(2018)]%
        {luciani2018details}
\bibfield{author}{\bibinfo{person}{Timothy Luciani}, \bibinfo{person}{Andrew
  Burks}, \bibinfo{person}{Cassiano Sugiyama}, \bibinfo{person}{Jonathan
  Komperda}, {and} \bibinfo{person}{G~Elisabeta Marai}.}
  \bibinfo{year}{2018}\natexlab{}.
\newblock \showarticletitle{Details-first, show context, overview last:
  supporting exploration of viscous fingers in large-scale ensemble
  simulations}.
\newblock \bibinfo{journal}{\emph{IEEE transactions on visualization and
  computer graphics}} \bibinfo{volume}{25}, \bibinfo{number}{1}
  (\bibinfo{year}{2018}), \bibinfo{pages}{1225--1235}.
\newblock


\bibitem[Marchionini(2006)]%
        {marchionini2006exploratory}
\bibfield{author}{\bibinfo{person}{Gary Marchionini}.}
  \bibinfo{year}{2006}\natexlab{}.
\newblock \showarticletitle{Exploratory search: from finding to understanding}.
\newblock \bibinfo{journal}{\emph{Commun. ACM}} \bibinfo{volume}{49},
  \bibinfo{number}{4} (\bibinfo{year}{2006}), \bibinfo{pages}{41--46}.
\newblock


\bibitem[Norman(2013)]%
        {norman2013design}
\bibfield{author}{\bibinfo{person}{Don Norman}.}
  \bibinfo{year}{2013}\natexlab{}.
\newblock \bibinfo{booktitle}{\emph{The design of everyday things: Revised and
  expanded edition}}.
\newblock \bibinfo{publisher}{Basic books}.
\newblock


\bibitem[Noy et~al\mbox{.}(2019)]%
        {noy2019industry}
\bibfield{author}{\bibinfo{person}{Natasha Noy}, \bibinfo{person}{Yuqing Gao},
  \bibinfo{person}{Anshu Jain}, \bibinfo{person}{Anant Narayanan},
  \bibinfo{person}{Alan Patterson}, {and} \bibinfo{person}{Jamie Taylor}.}
  \bibinfo{year}{2019}\natexlab{}.
\newblock \showarticletitle{Industry-scale knowledge graphs: lessons and
  challenges}.
\newblock \bibinfo{journal}{\emph{Commun. ACM}} \bibinfo{volume}{62},
  \bibinfo{number}{8} (\bibinfo{year}{2019}), \bibinfo{pages}{36--43}.
\newblock


\bibitem[O'Connor et~al\mbox{.}(2010)]%
        {o2010tweetmotif}
\bibfield{author}{\bibinfo{person}{Brendan O'Connor}, \bibinfo{person}{Michel
  Krieger}, {and} \bibinfo{person}{David Ahn}.}
  \bibinfo{year}{2010}\natexlab{}.
\newblock \showarticletitle{Tweetmotif: Exploratory search and topic
  summarization for twitter}. In \bibinfo{booktitle}{\emph{Proceedings of the
  International AAAI Conference on Web and Social Media}},
  Vol.~\bibinfo{volume}{4}.
\newblock


\bibitem[Palani et~al\mbox{.}(2021)]%
        {palani2021conotate}
\bibfield{author}{\bibinfo{person}{Srishti Palani}, \bibinfo{person}{Zijian
  Ding}, \bibinfo{person}{Austin Nyugen}, \bibinfo{person}{Andrew Chuang},
  \bibinfo{person}{Stephen Macneil}, {and} \bibinfo{person}{Steven~P. Dow}.}
  \bibinfo{year}{2021}\natexlab{}.
\newblock \showarticletitle{CoNotate: Suggesting Queries Based on Notes
  Promotes Knowledge Discover}. In \bibinfo{booktitle}{\emph{Proceedings of the
  SIGCHI conference on Human factors in computing systems}}.
\newblock


\bibitem[Peltonen et~al\mbox{.}(2017)]%
        {peltonen2017topic}
\bibfield{author}{\bibinfo{person}{Jaakko Peltonen}, \bibinfo{person}{Kseniia
  Belorustceva}, {and} \bibinfo{person}{Tuukka Ruotsalo}.}
  \bibinfo{year}{2017}\natexlab{}.
\newblock \showarticletitle{Topic-relevance map: Visualization for improving
  search result comprehension}. In \bibinfo{booktitle}{\emph{Proceedings of the
  22nd international conference on intelligent user interfaces}}.
  \bibinfo{pages}{611--622}.
\newblock


\bibitem[Radensky et~al\mbox{.}(2022)]%
        {radensky2022exploring}
\bibfield{author}{\bibinfo{person}{Marissa Radensky}, \bibinfo{person}{Doug
  Downey}, \bibinfo{person}{Kyle Lo}, \bibinfo{person}{Zoran Popovic}, {and}
  \bibinfo{person}{Daniel~S Weld}.} \bibinfo{year}{2022}\natexlab{}.
\newblock \showarticletitle{Exploring the Role of Local and Global Explanations
  in Recommender Systems}. In \bibinfo{booktitle}{\emph{CHI Conference on Human
  Factors in Computing Systems Extended Abstracts}}. \bibinfo{pages}{1--7}.
\newblock


\bibitem[Raskin(2000)]%
        {raskin2000humane}
\bibfield{author}{\bibinfo{person}{Jef Raskin}.}
  \bibinfo{year}{2000}\natexlab{}.
\newblock \bibinfo{booktitle}{\emph{The humane interface: new directions for
  designing interactive systems}}.
\newblock \bibinfo{publisher}{Addison-Wesley Professional}.
\newblock


\bibitem[Ribeiro et~al\mbox{.}(2016)]%
        {lime2016}
\bibfield{author}{\bibinfo{person}{Marco~Tulio Ribeiro},
  \bibinfo{person}{Sameer Singh}, {and} \bibinfo{person}{Carlos Guestrin}.}
  \bibinfo{year}{2016}\natexlab{}.
\newblock \showarticletitle{"Why Should I Trust You?": Explaining the
  Predictions of Any Classifier}. In \bibinfo{booktitle}{\emph{Proceedings of
  the 22nd ACM SIGKDD International Conference on Knowledge Discovery and Data
  Mining}} (San Francisco, California, USA) \emph{(\bibinfo{series}{KDD '16})}.
  \bibinfo{publisher}{Association for Computing Machinery},
  \bibinfo{address}{New York, NY, USA}, \bibinfo{pages}{1135–1144}.
\newblock
\showISBNx{9781450342322}
\urldef\tempurl%
\url{https://doi.org/10.1145/2939672.2939778}
\showDOI{\tempurl}


\bibitem[Ruotsalo et~al\mbox{.}(2013)]%
        {ruotsalo2013supporting}
\bibfield{author}{\bibinfo{person}{Tuukka Ruotsalo},
  \bibinfo{person}{Kumaripaba Athukorala}, \bibinfo{person}{Dorota
  G{\l}owacka}, \bibinfo{person}{Ksenia Konyushkova}, \bibinfo{person}{Antti
  Oulasvirta}, \bibinfo{person}{Samuli Kaipiainen}, \bibinfo{person}{Samuel
  Kaski}, {and} \bibinfo{person}{Giulio Jacucci}.}
  \bibinfo{year}{2013}\natexlab{}.
\newblock \showarticletitle{Supporting exploratory search tasks with
  interactive user modeling}.
\newblock \bibinfo{journal}{\emph{Proceedings of the American Society for
  Information Science and Technology}} \bibinfo{volume}{50},
  \bibinfo{number}{1} (\bibinfo{year}{2013}), \bibinfo{pages}{1--10}.
\newblock


\bibitem[Schwarz and Morris(2011)]%
        {schwarz2011augmenting}
\bibfield{author}{\bibinfo{person}{Julia Schwarz} {and}
  \bibinfo{person}{Meredith Morris}.} \bibinfo{year}{2011}\natexlab{}.
\newblock \showarticletitle{Augmenting web pages and search results to support
  credibility assessment}. In \bibinfo{booktitle}{\emph{Proceedings of the
  SIGCHI conference on human factors in computing systems}}.
  \bibinfo{pages}{1245--1254}.
\newblock


\bibitem[Sciascio et~al\mbox{.}(2019)]%
        {sciascio2019roadmap}
\bibfield{author}{\bibinfo{person}{Cecilia~Di Sciascio}, \bibinfo{person}{Peter
  Brusilovsky}, \bibinfo{person}{Christoph Trattner}, {and}
  \bibinfo{person}{Eduardo Veas}.} \bibinfo{year}{2019}\natexlab{}.
\newblock \showarticletitle{A Roadmap to User-Controllable Social Exploratory
  Search}.
\newblock \bibinfo{journal}{\emph{ACM Transactions on Interactive Intelligent
  Systems (TiiS)}} \bibinfo{volume}{10}, \bibinfo{number}{1}
  (\bibinfo{year}{2019}), \bibinfo{pages}{1--38}.
\newblock


\bibitem[Sciascio et~al\mbox{.}(2016)]%
        {di2016rank}
\bibfield{author}{\bibinfo{person}{Cecilia~Di Sciascio},
  \bibinfo{person}{Vedran Sabol}, {and} \bibinfo{person}{Eduardo~E Veas}.}
  \bibinfo{year}{2016}\natexlab{}.
\newblock \showarticletitle{Rank as you go: User-driven exploration of search
  results}. In \bibinfo{booktitle}{\emph{Proceedings of the 21st international
  conference on intelligent user interfaces}}. \bibinfo{pages}{118--129}.
\newblock


\bibitem[Sen et~al\mbox{.}(2006)]%
        {sen2006tagging}
\bibfield{author}{\bibinfo{person}{Shilad Sen}, \bibinfo{person}{Shyong~K Lam},
  \bibinfo{person}{Al~Mamunur Rashid}, \bibinfo{person}{Dan Cosley},
  \bibinfo{person}{Dan Frankowski}, \bibinfo{person}{Jeremy Osterhouse},
  \bibinfo{person}{F~Maxwell Harper}, {and} \bibinfo{person}{John Riedl}.}
  \bibinfo{year}{2006}\natexlab{}.
\newblock \showarticletitle{Tagging, communities, vocabulary, evolution}. In
  \bibinfo{booktitle}{\emph{Proceedings of the 2006 20th anniversary conference
  on Computer supported cooperative work}}. \bibinfo{pages}{181--190}.
\newblock


\bibitem[Shneiderman(2003)]%
        {shneiderman2003eyes}
\bibfield{author}{\bibinfo{person}{Ben Shneiderman}.}
  \bibinfo{year}{2003}\natexlab{}.
\newblock \showarticletitle{The eyes have it: A task by data type taxonomy for
  information visualizations}.
\newblock In \bibinfo{booktitle}{\emph{The craft of information
  visualization}}. \bibinfo{publisher}{Elsevier}, \bibinfo{pages}{364--371}.
\newblock


\bibitem[Suchanek et~al\mbox{.}(2007)]%
        {Suchanek2007YagoAC}
\bibfield{author}{\bibinfo{person}{Fabian~M. Suchanek},
  \bibinfo{person}{Gjergji Kasneci}, {and} \bibinfo{person}{Gerhard Weikum}.}
  \bibinfo{year}{2007}\natexlab{}.
\newblock \showarticletitle{Yago: a core of semantic knowledge}. In
  \bibinfo{booktitle}{\emph{WWW '07}}.
\newblock


\bibitem[Teevan et~al\mbox{.}(2005)]%
        {teevan2005personalizing}
\bibfield{author}{\bibinfo{person}{Jaime Teevan}, \bibinfo{person}{Susan~T
  Dumais}, {and} \bibinfo{person}{Eric Horvitz}.}
  \bibinfo{year}{2005}\natexlab{}.
\newblock \showarticletitle{Personalizing search via automated analysis of
  interests and activities}. In \bibinfo{booktitle}{\emph{Proceedings of the
  28th annual international ACM SIGIR conference on Research and development in
  information retrieval}}. \bibinfo{pages}{449--456}.
\newblock


\bibitem[Toch et~al\mbox{.}(2012)]%
        {toch2012personalization}
\bibfield{author}{\bibinfo{person}{Eran Toch}, \bibinfo{person}{Yang Wang},
  {and} \bibinfo{person}{Lorrie~Faith Cranor}.}
  \bibinfo{year}{2012}\natexlab{}.
\newblock \showarticletitle{Personalization and privacy: a survey of privacy
  risks and remedies in personalization-based systems}.
\newblock \bibinfo{journal}{\emph{User Modeling and User-Adapted Interaction}}
  \bibinfo{volume}{22}, \bibinfo{number}{1-2} (\bibinfo{year}{2012}),
  \bibinfo{pages}{203--220}.
\newblock


\bibitem[Valenzuela et~al\mbox{.}(2015)]%
        {valenzuela2015identifying}
\bibfield{author}{\bibinfo{person}{Marco Valenzuela}, \bibinfo{person}{Vu Ha},
  {and} \bibinfo{person}{Oren Etzioni}.} \bibinfo{year}{2015}\natexlab{}.
\newblock \showarticletitle{Identifying meaningful citations}. In
  \bibinfo{booktitle}{\emph{Workshops at the twenty-ninth AAAI conference on
  artificial intelligence}}.
\newblock


\bibitem[Vincent(2016)]%
        {vincent_2016}
\bibfield{author}{\bibinfo{person}{James Vincent}.}
  \bibinfo{year}{2016}\natexlab{}.
\newblock \bibinfo{title}{Apple boasts about sales; Google boasts about how
  good its AI is}.
\newblock
\newblock
\urldef\tempurl%
\url{https://www.theverge.com/2016/10/4/13122406/google-phone-event-stats}
\showURL{%
\tempurl}


\bibitem[White et~al\mbox{.}(2006)]%
        {white2006supporting}
\bibfield{author}{\bibinfo{person}{Ryen~W White}, \bibinfo{person}{Bill Kules},
  \bibinfo{person}{Steven~M Drucker}, {et~al\mbox{.}}}
  \bibinfo{year}{2006}\natexlab{}.
\newblock \showarticletitle{Supporting exploratory search, introduction,
  special issue, communications of the ACM}.
\newblock \bibinfo{journal}{\emph{Commun. ACM}} \bibinfo{volume}{49},
  \bibinfo{number}{4} (\bibinfo{year}{2006}), \bibinfo{pages}{36--39}.
\newblock


\bibitem[White and Roth(2009)]%
        {white2009exploratory}
\bibfield{author}{\bibinfo{person}{Ryen~W White} {and} \bibinfo{person}{Resa~A
  Roth}.} \bibinfo{year}{2009}\natexlab{}.
\newblock \showarticletitle{Exploratory search: Beyond the query-response
  paradigm}.
\newblock \bibinfo{journal}{\emph{Synthesis lectures on information concepts,
  retrieval, and services}} \bibinfo{volume}{1}, \bibinfo{number}{1}
  (\bibinfo{year}{2009}), \bibinfo{pages}{1--98}.
\newblock


\bibitem[Yee et~al\mbox{.}(2003)]%
        {yee2003faceted}
\bibfield{author}{\bibinfo{person}{Ka-Ping Yee}, \bibinfo{person}{Kirsten
  Swearingen}, \bibinfo{person}{Kevin Li}, {and} \bibinfo{person}{Marti
  Hearst}.} \bibinfo{year}{2003}\natexlab{}.
\newblock \showarticletitle{Faceted metadata for image search and browsing}. In
  \bibinfo{booktitle}{\emph{Proceedings of the SIGCHI conference on Human
  factors in computing systems}}. \bibinfo{pages}{401--408}.
\newblock


\end{thebibliography}

\end{document}